\documentclass[aps,prd,twocolumn,showpacs,superscriptaddress,floatfix]{revtex4-1}
\usepackage{graphicx}  
\usepackage{dcolumn}   
\usepackage{amssymb}   
\usepackage{amsmath}   
\usepackage{multirow}
\usepackage{hyperref}
\usepackage[caption=false]{subfig}
\usepackage[normalem]{ulem}
\usepackage{xcolor}

\usepackage{slashed}
\pdfoutput=1

\RequirePackage{xspace}

\def\MB        {MiniBooNE\xspace}
\def\DMS       {dark matter\xspace}
\def\NUANCE    {\textsc{NUANCE}\xspace}
\def\BdNMC     {\textsc{BdNMC}\xspace}

\def\DMP       {\ensuremath{\chi}\xspace}
\def\DMV       {\ensuremath{V}\xspace}
\def\EvisE {\ensuremath{E_\text{vis}^{e}}\xspace}
\def\EnuQE {\ensuremath{E_\nu^\text{QE}}\xspace}
\def\costhetae {\ensuremath{\cos\theta_{e}}\xspace}

\def\m         {\ensuremath{\rm \,m}\xspace}
\def\cm         {\ensuremath{\rm \,cm}\xspace}

\def\us       {\ensuremath{\mathrm{\,\mu s}}\xspace}
\def\ns       {\ensuremath{\mathrm{\,ns}}\xspace}
\def\GeV       {\ensuremath{\mathrm{\,GeV}}\xspace}
\def\MeV       {\ensuremath{\mathrm{\,MeV}}\xspace}
\def\keV       {\ensuremath{\mathrm{\,keV}}\xspace}
\def\GeVc      {\ensuremath{{\GeV\,c^{-1}}}\xspace}

\def\MeVcc      {\ensuremath{{\MeV\,c^{-2}}}\xspace}
\def\GeVcc      {\ensuremath{{\GeV\,c^{-2}}}\xspace}

\def\mmev      {\ensuremath{{\mathrm{\,m\,Me\kern -0.1em V^{-1}}}\xspace}}

\newcommand{\mDMV}{\ensuremath{m_{V}}\xspace}
\newcommand{\mDMP}{\ensuremath{m_{\chi}}\xspace}
\newcommand{\CCQE}[1][]{#1\ensuremath{\text{CCQE}}\xspace}
\newcommand{\NCE}[1][]{#1\ensuremath{\text{NCE}}\xspace}
\newcommand{\NCpi}[1][]{#1\ensuremath{\text{NC}{\pi}}\xspace}
\newcommand{\NCPi}[1][]{#1\ensuremath{\text{NC}{\pi^0}}\xspace}
\newcommand{\NCPip}{\ensuremath{\text{NC}\pi^\pm}\xspace}

\newcommand{\CCQENu}{\ensuremath{\CCQE_\nu}\xspace}
\newcommand{\NCENu}{\ensuremath{\NCE_\nu}\xspace}
\newcommand{\CCQENuBar}{\ensuremath{\CCQE_{\bar{\nu}}}\xspace}
\newcommand{\NCENuBar}{\ensuremath{\NCE_{\bar{\nu}}}\xspace}
\newcommand{\CCQEOff}{\ensuremath{\CCQE_\text{Off}}\xspace}
\newcommand{\NCEOff}{\ensuremath{\NCE_\text{Off}}\xspace}
\newcommand{\NCEOffTiming}{\ensuremath{\NCEOff^\text{Timing}}\xspace}
\newcommand{\NCPiNu}{\ensuremath{\NCPi_\nu}\xspace}
\newcommand{\NCPiNuBar}{\ensuremath{\NCPi_{\bar{\nu}}}\xspace}
\newcommand{\NCPiOff}{\ensuremath{\NCPi_\text{Off}}\xspace}
\newcommand{\NCPiOffTiming}{\ensuremath{\NCPiOff^\text{Timing}}\xspace}
\newcommand{\NCEE}[1][]{#1\ensuremath{\nu\text{-}e}\xspace}

\newcommand{\nuModeLong}{neutrino-mode\xspace}

\newcommand{\detBRB}{\ensuremath{\nu_\text{det}}\xspace}
\newcommand{\dirtBRB}{\ensuremath{\text{dirt}}\xspace}

\def\kineticMixing {\ensuremath{\epsilon}\xspace}
\def\darkFineStructure {\ensuremath{\alpha_{_D}}\xspace}
\def\gaugeCoupling {\ensuremath{g_{_D}}\xspace}
\def\gaugeCouplingB {\ensuremath{g_{_B}}\xspace}
\def\DMScale {\ensuremath{\kineticMixing^4\darkFineStructure}\xspace}
\def\Lagr      {\ensuremath{\mathcal{L}}\xspace}

\begin{document}

\widetext


\title{Dark matter search in nucleon, pion, and electron channels from a proton beam dump with \MB}
\newcommand{\alabama}{University of Alabama, Tuscaloosa, AL 35487}
\newcommand{\anl}{Argonne National Laboratory, Argonne, IL 60439}
\newcommand{\chicago}{University of Chicago, Chicago, IL, 60637}
\newcommand{\cincinnati}{University of Cincinnati, Cincinnati, OH 45221}
\newcommand{\columbia}{Columbia University, New York, NY 10027}
\newcommand{\ctpu}{Center for Theoretical Physics of the Universe, Institute for Basic Science (IBS), Daejeon, 34051, Korea}
\newcommand{\fermi}{Fermi National Accelerator Laboratory, Batavia, IL 60510}
\newcommand{\florida}{University of Florida, Gainesville, FL 32611}
\newcommand{\indiana}{Indiana University, Bloomington, IN 47405}
\newcommand{\lanl}{Los Alamos National Laboratory, Los Alamos, NM 87545}
\newcommand{\michigan}{University of Michigan, Ann Arbor, MI 48111}
\newcommand{\mexico}{Instituto de Ciencias Nucleares, Universidad Nacional Aut\'onoma de M\'exico, Mexico City 04510, Mexico}
\newcommand{\nmsu}{New Mexico State University, Las Cruces, NM 88003}
\newcommand{\pittsburgh}{University of Pittsburgh, Pittsburgh, PA 15260, USA}
\newcommand{\qmery}{Queen Mary University of London, London, E1 4NS, UK}
\newcommand{\smary}{Saint Mary's University of Minnesota, Winona, MN 55987}
\newcommand{\uta}{University of Texas (Arlington), Arlington, TX 76019}

\author{A.A.~Aguilar-Arevalo}\affiliation{\mexico}
\author{M.~Backfish}\affiliation{\fermi}
\author{A.~Bashyal}\affiliation{\uta}
\author{B.~Batell}\affiliation{\pittsburgh}
\author{B.C.~Brown}\affiliation{\fermi}
\author{R.~Carr}\affiliation{\columbia}
\author{A.~Chatterjee}\affiliation{\uta}
\author{R.L.~Cooper}\affiliation{\indiana}\affiliation{\nmsu}
\author{P.~deNiverville}\affiliation{\ctpu}
\author{R.~Dharmapalan}\affiliation{\anl}
\author{Z.~Djurcic}\affiliation{\anl}
\author{R.~Ford}\affiliation{\fermi}
\author{F.G.~Garcia}\affiliation{\fermi}
\author{G.T.~Garvey}\affiliation{\lanl}
\author{J.~Grange}\affiliation{\anl}\affiliation{\florida}
\author{J.A.~Green}\affiliation{\lanl}
\author{E.-C.~Huang}\affiliation{\lanl}
\author{W.~Huelsnitz}\affiliation{\lanl}
\author{I.L.~de Icaza Astiz}\affiliation{\mexico}
\author{G.~Karagiorgi}\affiliation{\columbia}
\author{T.~Katori}\affiliation{\qmery}
\author{W.~Ketchum}\affiliation{\lanl}
\author{T.~Kobilarcik}\affiliation{\fermi}
\author{Q.~Liu}\affiliation{\lanl}
\author{W.C.~Louis}\affiliation{\lanl}
\author{W.~Marsh}\affiliation{\fermi}
\author{C.D.~Moore}\affiliation{\fermi}
\author{G.B.~Mills}\thanks{Deceased}\affiliation{\lanl}
\author{J.~Mirabal}\affiliation{\lanl}
\author{P.~Nienaber}\affiliation{\smary}
\author{Z.~Pavlovic}\affiliation{\lanl}
\author{D.~Perevalov}\affiliation{\fermi}
\author{H.~Ray}\affiliation{\florida}
\author{B.P.~Roe}\affiliation{\michigan}
\author{M.H.~Shaevitz}\affiliation{\columbia}
\author{S.~Shahsavarani}\affiliation{\uta}
\author{I.~Stancu}\affiliation{\alabama}
\author{R.~Tayloe}\affiliation{\indiana}
\author{C.~Taylor}\affiliation{\lanl}
\author{R.T.~Thornton}\affiliation{\lanl}
\author{R.G.~Van de Water}\affiliation{\lanl}
\author{W. Wester}\affiliation{\fermi}
\author{D.H.~White}\thanks{Deceased}\affiliation{\lanl}
\author{J. Yu}\affiliation{\uta}
\collaboration{The MiniBooNE-DM Collaboration}\noaffiliation

\date{\today}

\begin{abstract}
A search for sub-GeV dark matter produced from collisions of the Fermilab 8\GeV Booster protons 
with a steel beam dump was performed by the MiniBooNE-DM Collaboration using data from 
$1.86 \times 10^{20}$ protons on target in a dedicated run. 
The MiniBooNE detector, consisting of 818 tons of mineral oil and located 490 meters downstream 
of the beam dump, is sensitive to a variety of dark matter initiated scattering reactions. 
Three dark matter interactions are considered for this analysis: elastic scattering off nucleons, inelastic 
neutral pion production, and elastic scattering off electrons. 
Multiple data sets were used to constrain flux and systematic errors, and time-of-flight information was 
employed to increase sensitivity to higher dark matter masses. 
No excess from the background predictions was observed, and 90$\%$ confidence level limits were set 
on the vector portal and leptophobic dark matter models. 
New parameter space is excluded in the vector portal dark matter model with a dark matter mass between 
5 and 50\MeVcc. 
The reduced neutrino flux allowed to test if the MiniBooNE neutrino excess scales with the production of neutrinos. 
No excess of neutrino oscillation events were measured ruling out models that scale solely by number of protons on target independent of 
beam configuration  at 4.6$\sigma$. 
\end{abstract}

\maketitle


\section{\label{sec:intro}Introduction}
A wide variety of astrophysical and cosmological observations present strong evidence for the existence of 
dark matter, and a diverse experimental program has developed over the past few decades to search for its 
nongravitational interactions.
In the context of the popular weak scale ``WIMP'' dark matter scenarios, impressive coverage has been achieved 
through several experimental and observational approaches, including direct searches for dark matter scattering 
with nuclei, indirect searches for dark matter annihilation in the Galaxy and beyond, and high-energy collider 
searches for missing energy.
However, these traditional search strategies are often less sensitive to light dark matter candidates with masses
 below a few \GeVcc, and it is thus important to consider alternative experimental approaches to dark matter detection in this regime.

In this context, there has been a growing realization that fixed-target experiments can provide significant and 
complementary sensitivity to sub-GeV dark matter that couples to ordinary matter through a light 
mediator~\cite{Batell:2009di,deNiverville:2011it,deNiverville:2012ij,Dharmapalan:2012xp,Izaguirre:2013uxa,Izaguirre:2014bca,
Izaguirre:2014dua,Batell:2014yra,Batell:2014mga,Dobrescu:2014ita,Soper:2014ska,Kahn:2014sra,Izaguirre:2015yja,deNiverville:2015mwa,
Izaguirre:2015pva,Coloma:2015pih,Izaguirre:2017bqb,Frugiuele:2017zvx,Magill:2018tbb}.
In this approach, a relativistic flux of dark matter particles is produced out of proton (or electron) collisions 
with a fixed target, followed by the detection of the dark matter through its scattering in a detector placed 
downstream of the target.
This approach was successfully employed at Fermilab with the \MB detector, setting new limits on sub-GeV 
dark matter in the neutral-current (quasi )elastic nucleon scattering with no pion in the final state (\NCE)~\cite{Aguilar-Arevalo:2017mqx}.

\begin{figure}[hbtp!]
  \includegraphics[width=0.5\textwidth]{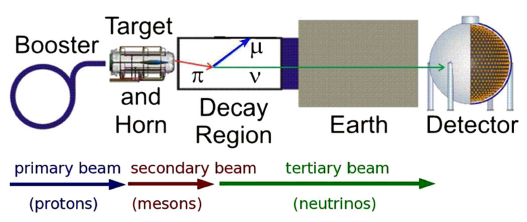}
\caption{\label{fig:fig1}The production of neutrinos in the Booster Neutrino Beamline in on-target 
running~\cite{AguilarArevalo:2010zc}.}
\end{figure}

The \MB experiment was designed to study short-baseline neutrino oscillations~\cite{Aguilar-Arevalo:2013pmq}.
In the normal neutrino or antineutrino running modes, charged pions $\pi^\pm$ are produced in the collisions 
of the proton beam with a beryllium target and subsequently decay in flight to neutrinos in the decay volume 
immediately following the target, as shown in Fig~\ref{fig:fig1}.
This results in a large flux of neutrinos at the \MB detector, which is a background to the \DMS neutral-current-like 
scattering signature.
Instead, in the beam-dump running mode, the proton beam is steered past the beryllium target and directed onto 
the steel absorber at the end of the decay volume, which significantly reduces the neutrino flux and increases 
sensitivity to a potential dark matter signal.
A dedicated run in beam-dump mode was carried out from November 2013 to September 2014 collecting 
$1.86\times10^{20}$\, protons on target (POT).
Besides the capability of running in beam-dump mode, \MB has several advantages which make this search feasible, 
including a detailed understanding of detector response and standard background processes gained through over a 
decade of operation, and robust and well-tested particle identification techniques.

The results presented here improve upon those in Ref.~\cite{Aguilar-Arevalo:2017mqx} by including two additional 
dark matter interaction channels in two separate analyses. 
The first was a combined \NCE and neutral-current pion production through delta resonant decay (\NCpi) fit to search 
for dark matter interaction with nucleons, and the second was dark matter elastically scattering off electrons. 
A ``time-of-flight'' observable was also added to both analyses to increase sensitivity to heavier dark matter masses. 
No significant excess is observed in either analysis, and 90\% confidence level limits are derived for vector portal 
and leptophobic dark matter models.
\MB excludes new parameter space in the vector portal dark matter model.

Results from applying the neutrino oscillation cuts are also presented. 
With the reduction of the neutrino flux, a test was preformed to determine if the neutrino oscillation 
excess~\cite{Aguilar-Arevalo:2013pmq,Aguilar-Arevalo:2018gpe} comes from a process that scales with 
neutrino production or a process that would scale solely on the number of POT.  

The following section provides an overview of the theoretical aspects of sub-GeV dark matter.
 Following this, Sec.~\ref{sec:beamline} reviews the Booster Neutrino Beamline (BNB), where the neutrino flux  
 (in beam-dump mode) is given, and the ``time-of-flight'' measurement is discussed.
 In Sec.~\ref{sec:detector} the \MB detector and simulations are reviewed.
 Section~\ref{sec:distributions} we present the event distributions, and describe backgrounds, systematics, and fit methodology.
 Finally, the dark matter results are presented in Sec.~\ref{sec:modelDep}, and a discussion of the implications for both the 
 dark matter and neutrino oscillation searches is given in Sec.~\ref{sec:results}.


\section{\label{sec:theory}Theory of sub-GeV Dark Matter}
Light dark matter \DMP with a mass below 1\GeVcc and coupled to ordinary matter through a light mediator particle is a 
viable and theoretically well-motivated possibility.
 While it is possible that \DMP exists at this scale in isolation, on general grounds one may expect a larger ``dark sector'' 
 of states.
 One or more of these additional states may mediate interactions to the Standard Model (SM) and may also play a role in 
 the cosmological production of \DMS, allowing for the correct relic abundance through the standard thermal 
 freeze-out mechanism.

The simplest dark sector scenario of this type is known as {\it vector portal dark matter}, in which the interactions of \DMP 
are mediated by a new dark $U(1)$ gauge boson $\DMV_\mu$ that kinetically mixes with the ordinary 
photon~\cite{Boehm:2003hm,Fayet:2004bw,Pospelov:2007mp,ArkaniHamed:2008qn,Holdom:1985ag,deNiverville:2016rqh}. 
 In such a model, there are four parameters that govern the properties of dark matter: the dark matter mass \mDMP, the 
 dark photon mass \mDMV, the kinetic mixing angle \kineticMixing, and the dark gauge coupling \gaugeCoupling. 
Equation~(\ref{eq:eq1}) gives the Lagrangian $\Lagr_V$ that is added to the SM Lagrangian: 
\begin{equation}
  \Lagr_V =\Lagr_\DMP-\frac{1}{4}V_{\mu\nu} V^{\mu\nu}+\frac{1}{2}\mDMV^2\DMV_{\mu}\DMV^{\mu}-
  \frac{\kineticMixing}{2}F_{\mu\nu}V^{\mu\nu},\label{eq:eq1}
\end{equation}
where
\[
\Lagr_\DMP = 
\left\{
  \begin{array}{l l}
    i\overline{\DMP}\slashed{D}\DMP - \mDMP\overline{\DMP}\DMP & \quad\text{Dirac fermion},\\
    \left|D_\mu\DMP\right|^2 - \mDMP^2\left|\DMP\right|^2 & \quad\text{Complex scalar},
  \end{array}
  \right.
\]
and $D_\mu = \partial_\mu-i\gaugeCoupling \DMV_{\mu}$ with the dark matter charge equal to one. 
The interactions above lead to efficient dark matter annihilation to light SM particles such that the observed dark matter 
abundance can be explained for certain values of the model parameters. 
Furthermore, if the dark matter is a complex scalar the annihilation occurs in the $p$-wave and is velocity 
suppressed~\cite{deNiverville:2011it}, 
evading otherwise strong constraints from the cosmic microwave background~\cite{Ade:2015xua}. 
For this reason, the \DMS particle is assumed to be a complex scalar in this work.

\begin{figure}[htbp]
  \centering
  \subfloat[Meson Decay]{\includegraphics[width=0.20\textwidth]{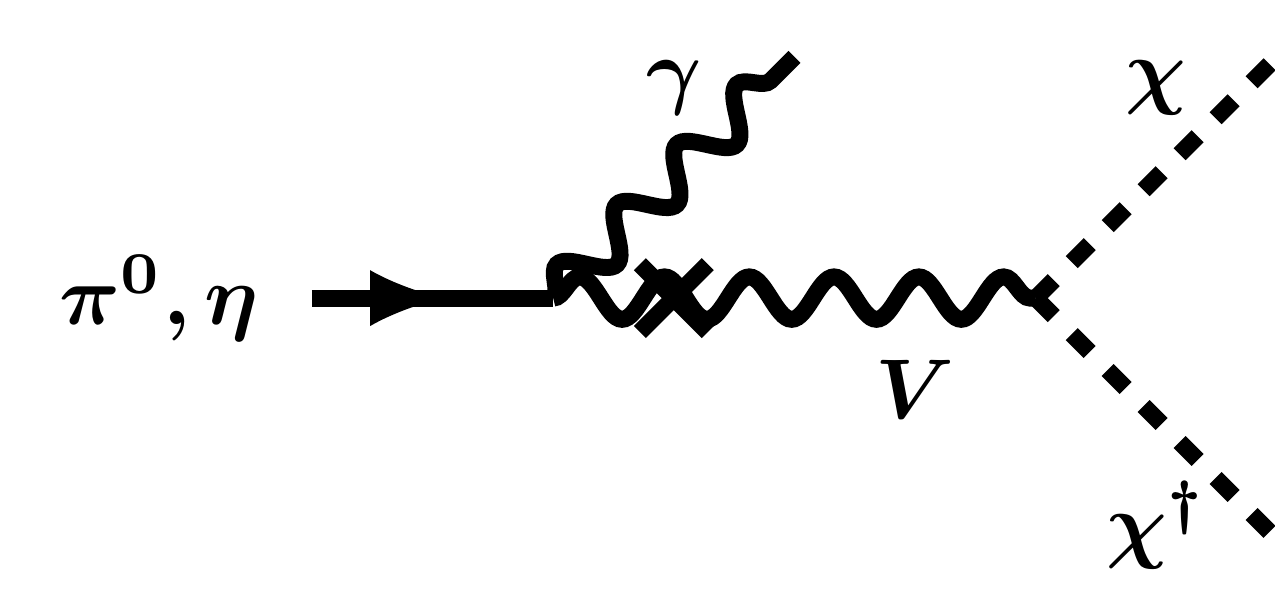}}\qquad%
  \subfloat[Proton Bremsstrahlung + Vector-Mixing]{\includegraphics[width=0.20\textwidth]{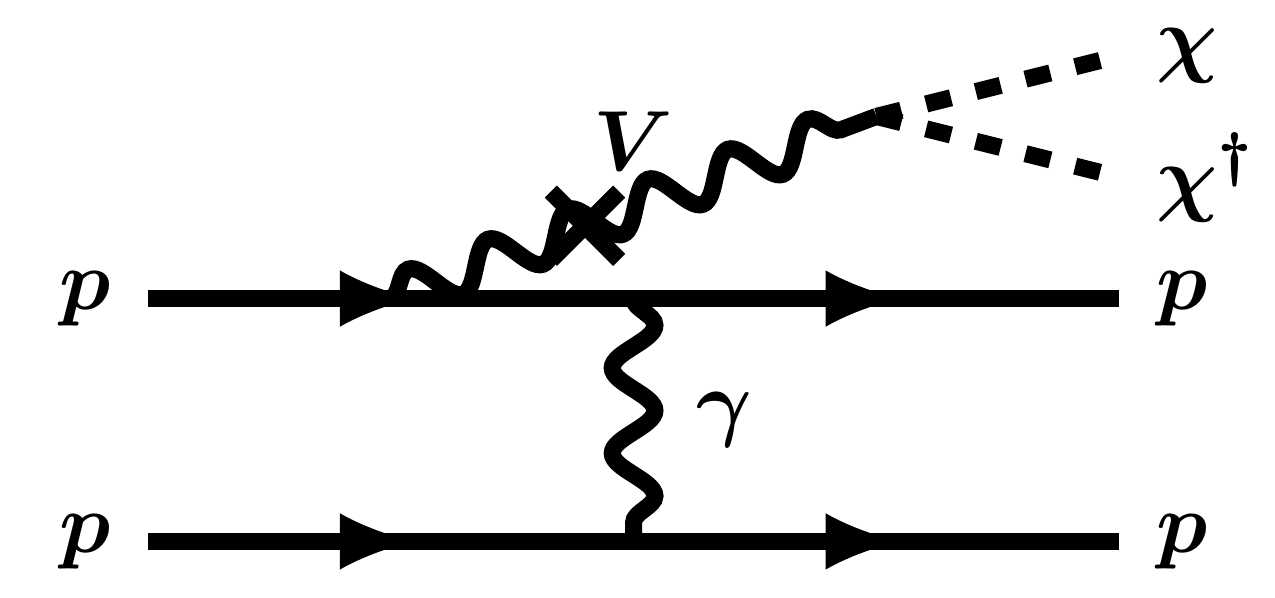}}
  \caption{\label{fig:fig2}Feynman diagrams for the production channels relevant for the \MB dark matter 
  search~\cite{Aguilar-Arevalo:2017mqx}.}
\end{figure}

The BNB is able to produce dark matter through several mechanisms, illustrated in Fig.~\ref{fig:fig2}.
They are 
(i) decay of secondary $\pi^0$ or $\eta$ mesons, and
(ii) proton bremsstrahlung plus vector-meson mixing.
  Note that in all cases, the production rate scales as $\kineticMixing^2$ provided \DMV can decay to two
  on-shell \DMP.
  On-shell decay is defined by  $\mDMV > 2 \mDMP$, and is known as the invisible decay mode.
     
\begin{figure}[htbp]
  \centering
  \subfloat[Free Protons or Electrons]{\includegraphics[width=0.20\textwidth]{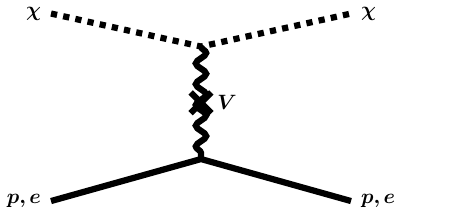}}\qquad%
  \subfloat[Bound Nucleons]{\includegraphics[width=0.20\textwidth]{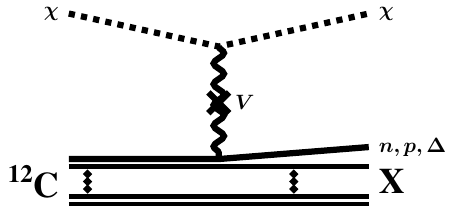}}
  \caption{\label{fig:fig3}Feynman diagrams for the dark matter interactions with nucleons and electrons 
  in \MB. The $\Delta$, in the bound-nucleon case, would be observed by its decay products, a pion and a nucleon.}
\end{figure}

  Once the dark matter is produced by one of these mechanisms, it can scatter with nucleons or electrons through a 
  neutral-current channel in the detector via $\DMV_\mu$ boson exchange, as depicted in Fig.~\ref{fig:fig3}.
  The scattering rate scales as $\kineticMixing^2 \darkFineStructure$, where $\darkFineStructure = \gaugeCoupling^2/4\pi$.
  The accelerator-produced \DMS event rate in \MB scales as  $\kineticMixing^4\darkFineStructure$ for on-shell decays in this model. 

Another potential dark sector scenario amenable to the \MB search is {\it leptophobic dark matter}%
~\cite{Fayet:1990wx,Fayet:2016nyc,Batell:2014yra,Dobrescu:2014ita,Soper:2014ska}, in which the mediator \DMV couples dominantly to quarks and 
not leptons. 
For illustration, a simplified scenario is presented in which a vector mediator couples to the baryon number current, with 
the Lagrangian given in Eq.~(\ref{eq:eq2}):
\begin{equation}
  \Lagr_B=\Lagr_V + \gaugeCouplingB \DMV_\mu J_B^\mu + \cdots,\label{eq:eq2}
\end{equation}
where
\[
  J_\mu^B = \frac{1}{3}\displaystyle\sum_i\bar{q}_i\gamma_\mu q_i\text{    },
\]
is the sum over all quark species, and $\Lagr_V$ [Eq.~(\ref{eq1})] is dependent on the baryon gauge coupling 
$\gaugeCouplingB$ ($\gaugeCoupling$ is replaced by $\gaugeCouplingB$). 
The limit $\kineticMixing e \ll g_B$ gives the leptophobic dark matter scenario. 
Three parameters will be considered in the interpretation of the presented results: the dark matter mass $\mDMP$, 
the leptophobic vector mediator mass $\mDMV$, and the coupling $\alpha_{_B} = \gaugeCouplingB^2/4\pi$. 
Consideration of the dark matter production and scattering rates leads to the conclusion that the event rates scale 
as $\alpha_{_B}^3$ for on-shell decays. 

It turns out to be challenging to construct a phenomenologically viable UV completion of the leptophobic model with 
large mediator couplings to the SM. 
Among other challenges, significant constraints arise as a consequence of the 
anomalous nature of the vector mediator in the case at hand~\cite{Dobrescu:2014fca,Dror:2017ehi,Dror:2017nsg}, which will provide 
stronger constraints than the \MB dark matter search in most UV completions of the model. 
Nevertheless, the \MB limits presented here are likely to be of value in certain leptophobic scenarios, e.g., 
those involving leptophobic scalar mediators. 

As we are discussing new light degrees of freedom at the (sub-) GeV scale, a variety of constraints from past 
experiments must be considered. 
The strongest constraints on the scenarios discussed above arise from fixed-target/beam-dump experiments, 
medium-energy $e^+ e^-$ colliders, and meson decays. 
These constraints were described in detail in 
Refs.~\cite{Batell:2014mga,Banerjee:2017hhz,Essig:2017kqs,deNiverville:2016rqh,Lees:2014xha} for the vector 
portal model, and in Refs.~\cite{Batell:2014yra,Dror:2017ehi,Dror:2017nsg} for the leptophobic model.


\section{\label{sec:beamline}Booster Neutrino Beamline}
The Fermilab Booster delivers 8 GeV (kinetic energy) protons to the BNB target hall.
As shown in Fig.~\ref{fig:fig1}, when running in on-target mode  a secondary beam of mesons 
is produced that travel through the air-filled decay pipe and decay in flight to produce neutrinos which then 
travel and interact in the \MB detector.
The intensity of the proton beam can range from $1\times10^{12}$ protons per pulse (ppp) to 
$5\times10^{12}$ ppp.

\begin{figure}[htbp!]
\includegraphics[width=0.5\textwidth]{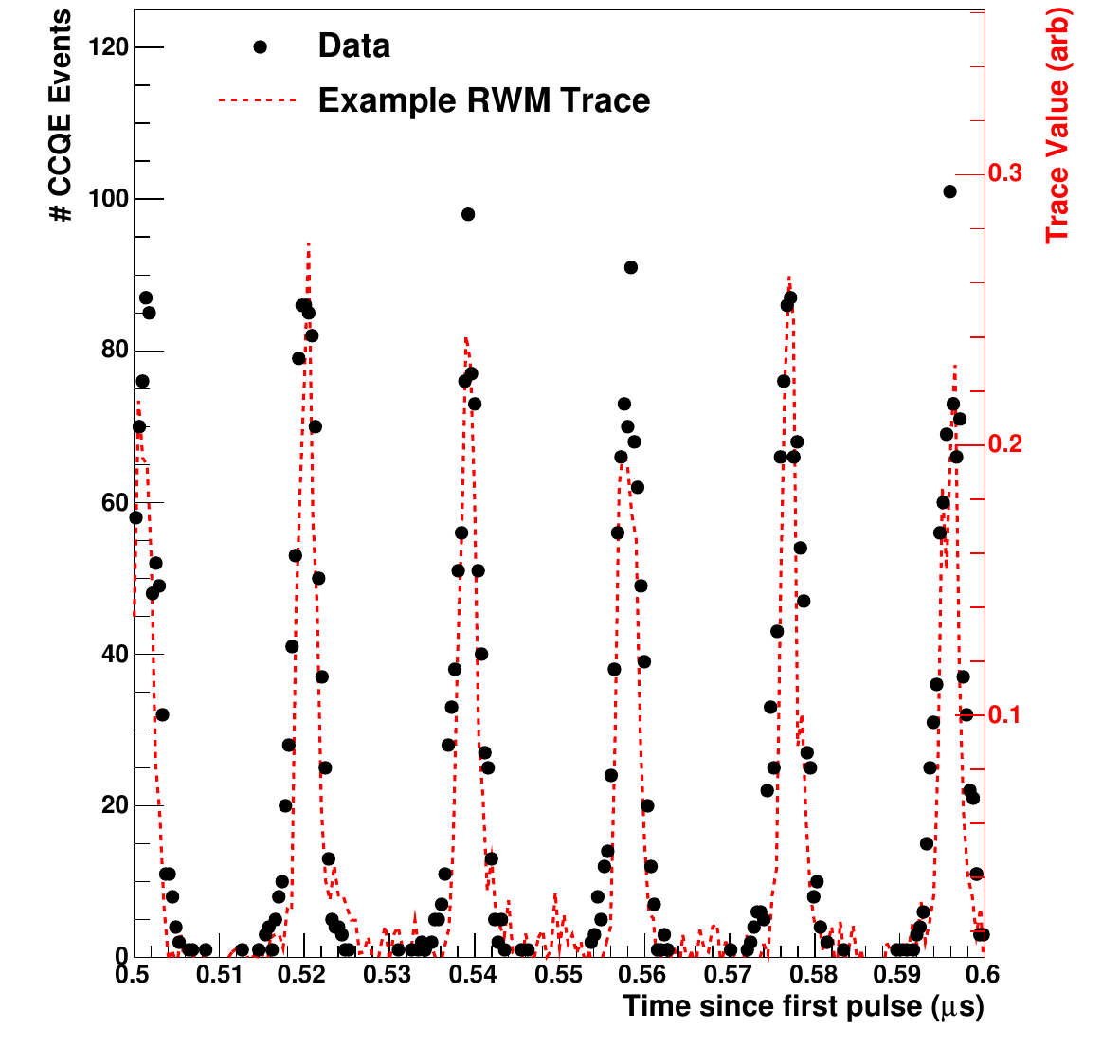}
\caption{\label{fig:fig4}Zoomed-in example of the BNB pulse microstructure as measured by the RWM. 
The data points come from  neutrino-mode $\nu_\mu$ charged-current interactions in the 
\MB detector during 2015--2016. The example RWM trace is plotted by the readout value of the trace.}
\end{figure}

Each pulse has a 53\,MHz microstructure that is composed of 82 bunches, and
each bunch has a full width half maximum of 2\ns.
Figure~\ref{fig:fig4} overlays an example trace of the BNB pulse microstructure, with an arbitrary offset with 
neutrino mode $\nu_\mu$ charged-current quasielastic (CCQE) interactions in the \MB detector; 
see Sec.~\ref{sec:distributions} for definition.
The trace and the CCQE data shapes are in good agreement.

\begin{figure}[hbtp!]
  \includegraphics[width=0.5\textwidth]{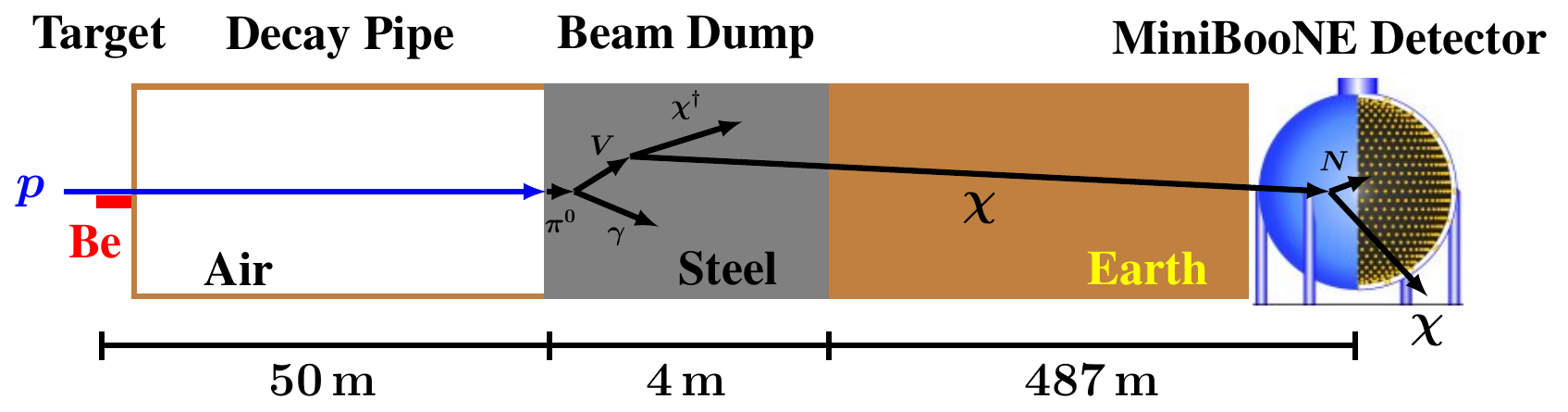}
\caption{\label{fig:fig5}The production of dark matter in off-target running~\cite{Aguilar-Arevalo:2017mqx}.}
\end{figure}

Neutrinos are a background for the \DMS search. 
To reduce the neutrino production coming from the BNB, the primary proton beam was steered above the beryllium target, 
and into a cooling air gap (which is inside the neck of the aluminum horn). 
After leaving the horn the protons enter the air-filled decay pipe, and finally 
reach the beam dump located 50\m downstream of the target location, as illustrated in Fig.~\ref{fig:fig5}. 
Running in this mode reduces the number of charged mesons that are generated in the thin beryllium target. 

\begin{figure}[hbtp!]
  \includegraphics[width=0.5\textwidth]{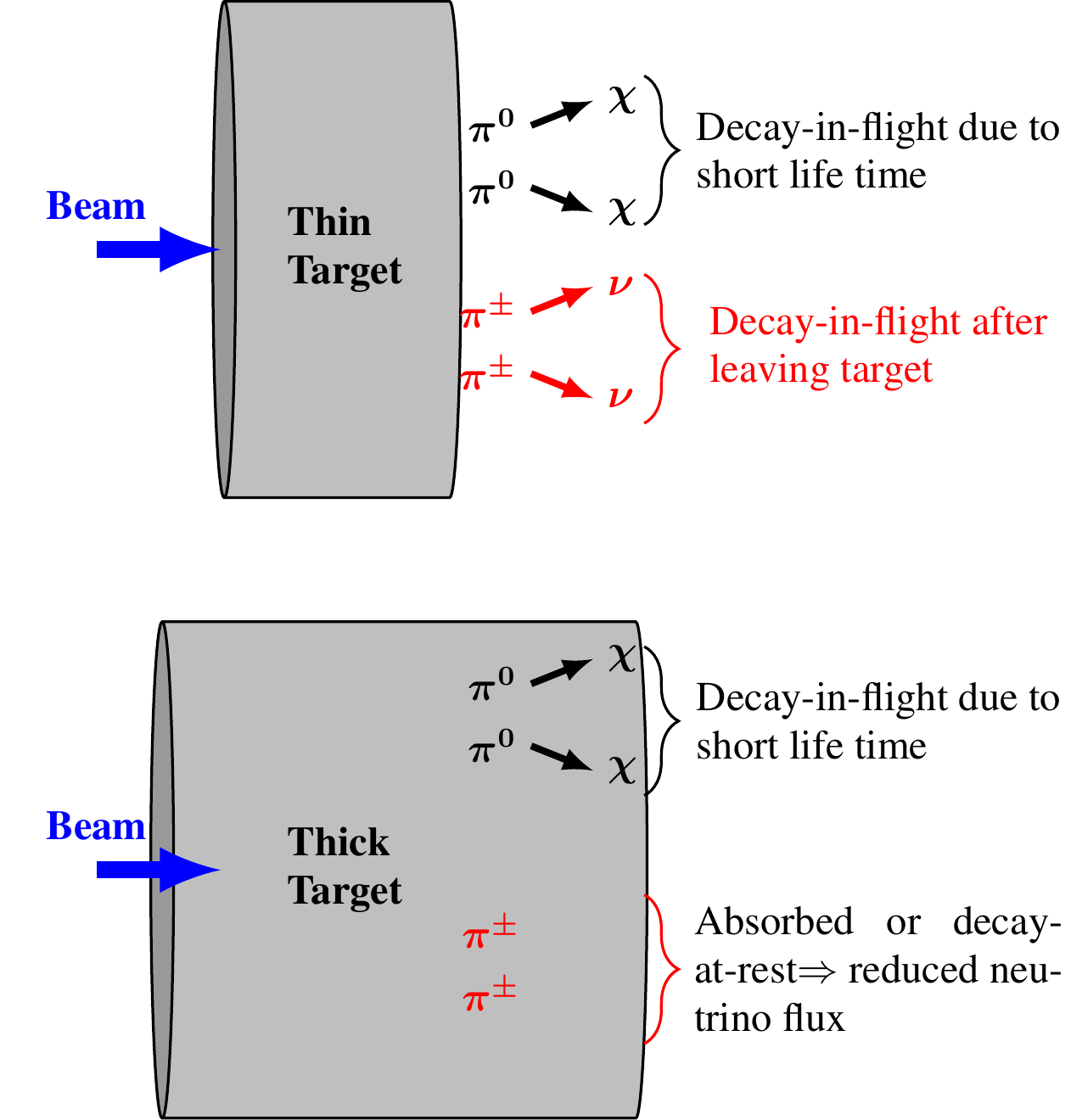}
\caption{\label{fig:fig6} (Top) Production of \DMS and neutrinos when the beam hits a thin target. 
(Bottom) The production of \DMS and suppression of neutrino generation when the beam hits a thick target.}
\end{figure}

The charged mesons that are produced in a thin target will escape and produce decay-in-flight neutrinos, while 
within the beam dump, the charged mesons are absorbed or decay at rest within a few radiation lengths, as 
illustrated in Fig.~\ref{fig:fig6}. 
This is in comparison with neutral mesons that will decay in flight due to their short lifetimes.
The neutral mesons could decay into a dark photon which would then decay into two dark matter particles, as shown schematically 
in Fig.~\ref{fig:fig5}.
The horn was turned off during this run so no charged particles generated would be (de)focused.
For the rest of this paper, this mode of running will be denoted as off-target, since the beryllium target and horn 
were not removed from the beam line.

The decay pipe and beam dump are buried in crushed aggregate.
There is a metal end cap at the downstream end of the decay pipe which prevents aggregate from entering the pipe.
The beam dump consists of 104 inches of steel followed by 36 inches of concrete and another 26 inches of steel 
in the beam direction.  
A detailed study of the neutrino flux coming from the BNB in on-target mode seen in the \MB detector using the 
\textsc{GEANT4}~\cite{Agostinelli:2002hh} simulation package \textsc{BooNEG4Beam} can be found in Ref.~\cite{AguilarArevalo:2008yp}. 
On-target running consisted of neutrino, and antineutrino modes.
The simulations were updated to study the off-target beam configuration and are described below.

\subsection{\label{subsec:flux} Beam off-target BNB simulation}
\begin{figure}[htbp]
  \includegraphics[width=0.5\textwidth]{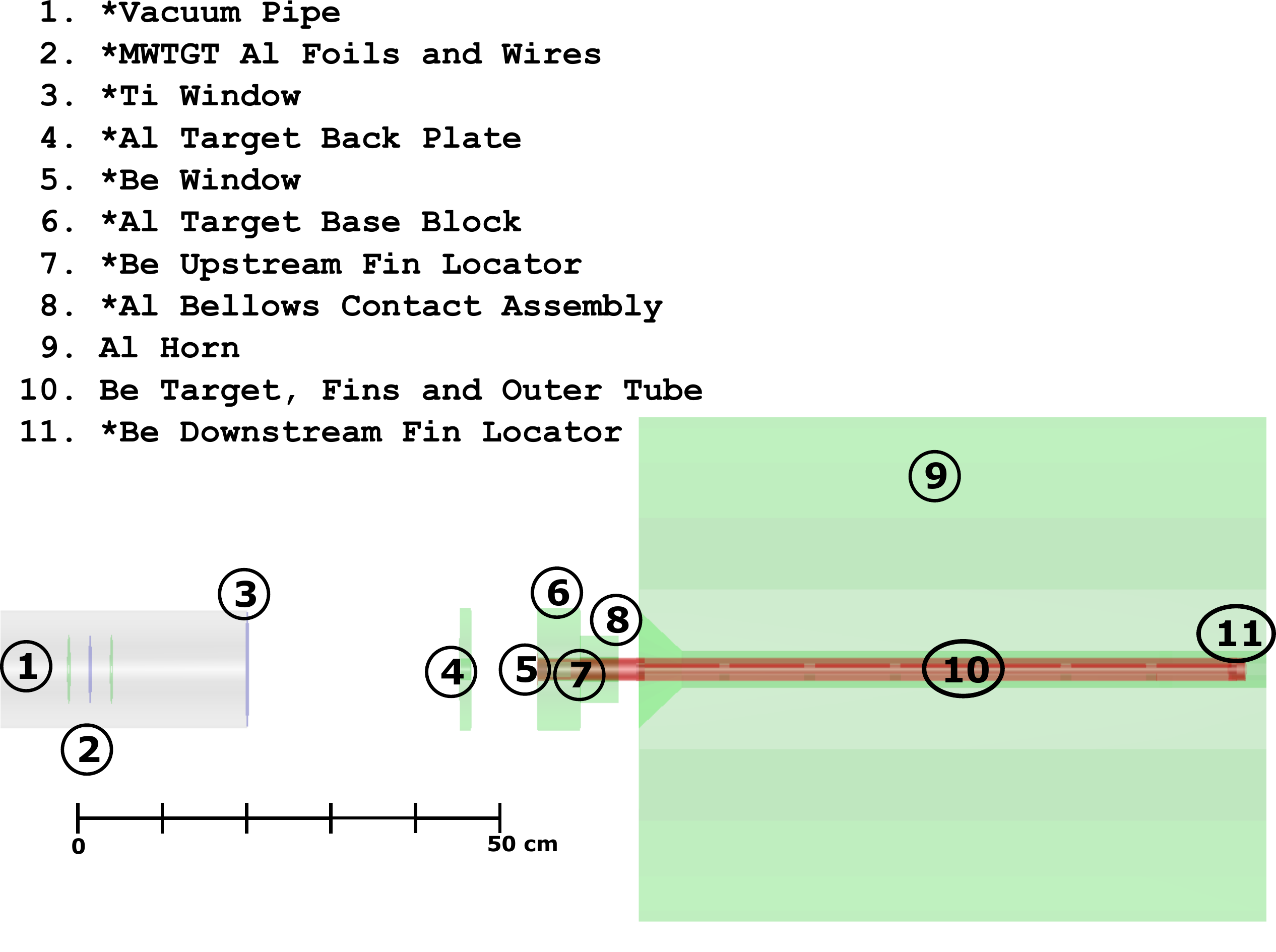}
  \caption{\label{fig:fig7}The simulated geometry around the target. 
  Those listed with an asterisk were added for the off-target simulation. 
  The added materials change the \nuModeLong flux by less than a percent.}
\end{figure}

\textsc{BooNEG4Beam} was updated to include materials in the beam line that would have changed the \nuModeLong 
flux $\Phi_\nu$ by less than a percent but are important for the off-target beam configuration.
Figure~\ref{fig:fig7} shows a schematic of the beam-line geometry around the target, pointing out 
the materials that were added.
An aluminum window at the end of the horn and a steel end cap with a small gap of air between the end of the 
beam pipe and the steel beam dump were also added. Except for the windows and the end cap, the other materials 
that were added are hollow around the beam center, and do not add to the primary meson production during on-target running.
The starting beam parameters for the off-target simulations were chosen by {\it in situ} measurements from two multiwire 
planes, about one meter apart and about four meters upstream of the target.

\begin{figure}[htbp!]
\includegraphics[width=0.5\textwidth]{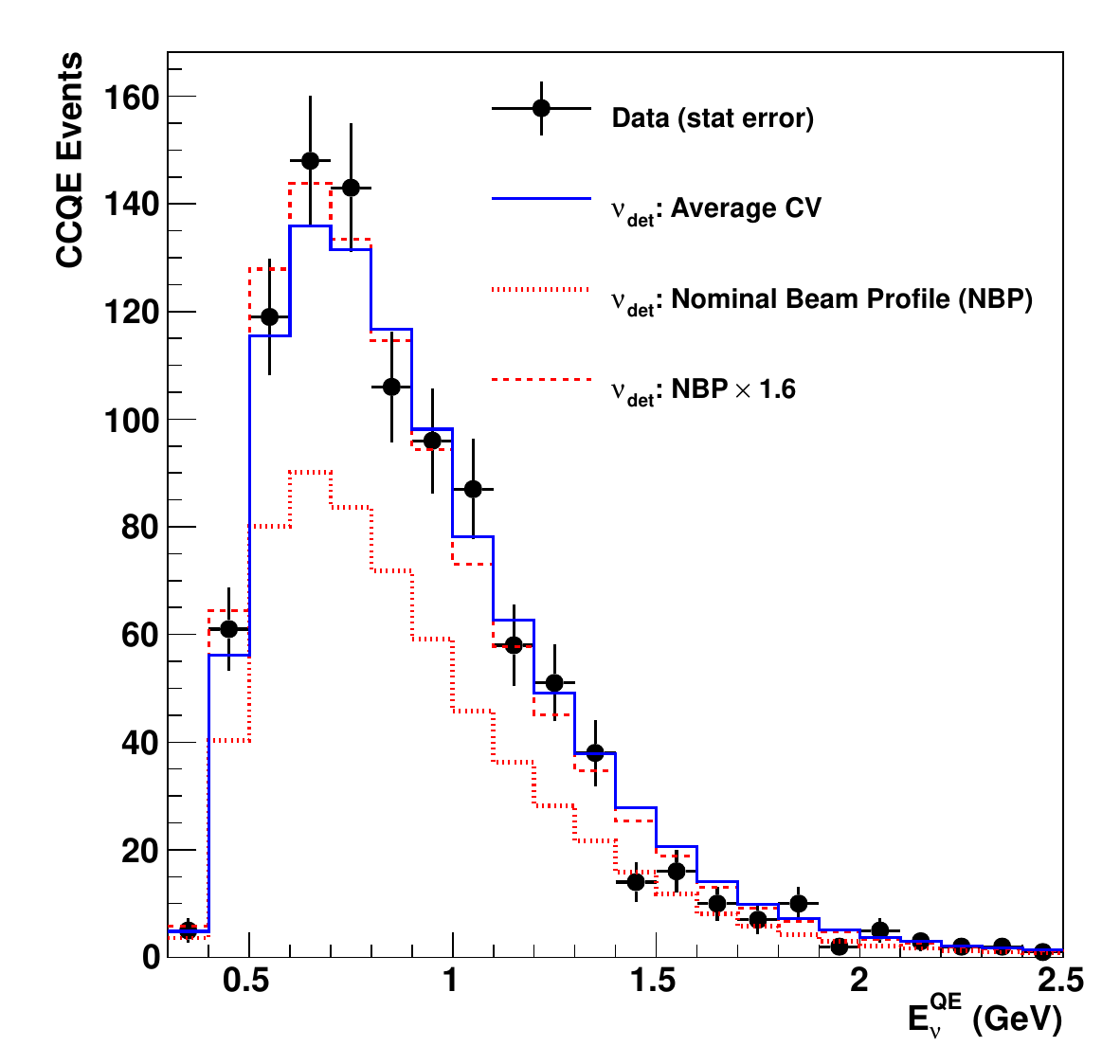}
\caption{\label{fig:fig8}Comparing \CCQE data in off-target mode to three different Monte Carlo predictions for 
neutrinos interacting in the detector (\detBRB). The dotted line is the output of the nominal off-target beam profile, the dashed line is the 
nominal profile scale by 1.6, and the solid line is the average of the scrapings (Average CV) used as the final 
$\Phi_\text{Off}$~\cite{Thornton:2017etu}. \EnuQE is defined by Eq.~\ref{eq:eq5}.}
\end{figure}

The \DMS model does not have a charged-current interaction component in its simplest form resulting in the assumption 
that the \CCQE signature in \MB (see Sec.~\ref{sec:distributions}) does not have a \DMS signal component. 
The \CCQE distribution was used to check the simulated off-target flux $\Phi_\text{Off}$.
The nominal off-target beam parameters and geometry produced 60\% less \CCQE events than measured, as shown 
in Fig.~\ref{fig:fig8}.

\begin{figure}[htbp!]
\includegraphics[width=0.5\textwidth]{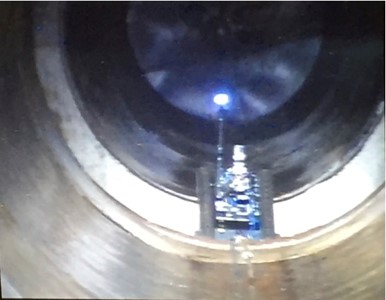}
\caption{\label{fig:fig9}Picture of FRED at the 25-\m absorber.}
\end{figure}

In August of 2015 a remote-controlled robotic vehicle was employed to survey the region between the target horn and 
the end of the decay pipe. 
The objective of the survey was to do a visual inspection of the space where the proton beam traveled in the decay pipe 
during off-target mode to determine if anything was causing the increase of \CCQE events. 
The Finding Radiation Evidence in the Decay pipe (FRED) robot was equipped with a Hall probe to measure any stray 
magnetic fields that could affect the beam direction, and a camera. 
See Fig.~\ref{fig:fig9} for a picture of FRED under the 25-m absorber.
The survey found that the magnetic field was within previous expectations and the space was clear of any unexpected 
debris or obstruction. 
The conclusion was that nothing in the decay pipe was causing the increase in the \CCQE rate.

\begin{figure}[htbp]
  \includegraphics[width=0.5\textwidth]{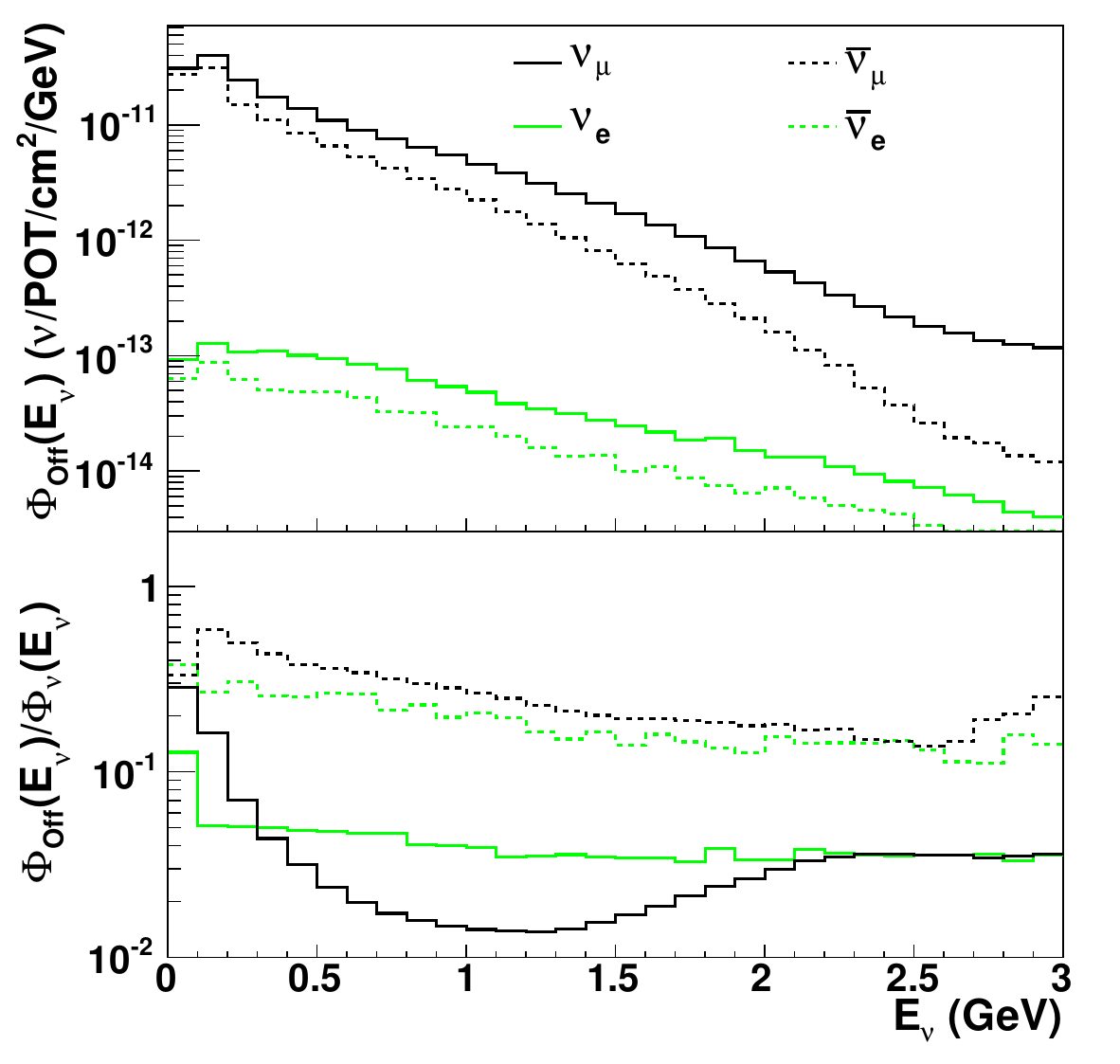}
  \caption{\label{fig:fig10}(Top) The off-target neutrino flux seen by the \MB detector. 
  (Bottom) The off-target/neutrino flux ratio~\cite{Aguilar-Arevalo:2017mqx}.}
\end{figure}

A simulation study was able to account for the increased rate by moving the primary beam angles within 2$\sigma$ of 
their uncertainties~\cite{Thornton:2017etu}. 
These small movements caused the tails of the beam to scrape the beryllium target downstream of the $90^\circ$ beam-loss monitor. 
The same study showed, with the low-statistics off-target data, that no distinction could be made between the different scrapings. 
An average of four potential scraping scenarios produced the needed increase in the number of \CCQE events. 
The average is defined as $\Phi_\text{Off}$, as shown in Fig.~\ref{fig:fig10}. 

Uncertainty in $\Phi_\text{Off}$ was determined by 1$\sigma$ excursions around the nominal beam profile scaled by 
1.6 so the central value number of \CCQE events matches data, and the four potential scraping scenarios that were 
averaged to generate $\Phi_\text{Off}$. 
The integrated $\Phi_\text{Off}$ with a neutrino energy $E_\nu$ between 0.2 and 3\GeV is 
$\left(1.9\pm1.1\right)\times 10^{-11}\,\nu/\mathrm{POT}/\mathrm{cm^{2}}$ with a mean energy of 660\MeV. 
The large uncertainty on $\Phi_\text{Off}$ comes from not knowing which scraping scenario is physically happening. 
Comparing this to the integrated $\Phi_\nu$ of $5.0\times10^{-10}\,\nu/\mathrm{POT}/\mathrm{cm^{2}}$ with a mean 
energy of 830\MeV gives a flux reduction factor of 27. 
The reduction factor as a function of $E_\nu$ and species is shown in Fig.~\ref{fig:fig10}. 
The combination of the flux reduction and the softer spectrum, which has lower neutrino cross sections in the detector, 
results in an event-rate reduction by a factor of 48 in both \CCQE and \NCE interactions (cuts given in Table~\ref{tab:tab3}). 
 
The breakdown of the integrated $\Phi_\text{Off}$ for the different neutrino species is given in Table~\ref{tab:tab1}.
\begin{table}[htbp!]
  \caption{\label{tab:tab1}Beam off-target profile systematic percent error independent of energy 
  for the various neutrino types, including correlations. $\Phi_\text{Off}$ integrated over $0.2 < E_\nu < 3\GeV$ }
  \begin{ruledtabular}
  \begin{tabular}{rrr}
    Neutrino species   & \multicolumn{1}{c}{$\Phi_\text{Off} \left(\nu/POT/cm^2\right)$} & \% of total\\
    Total           & $\left(1.9\pm1.1\right)\times10^{-11}$\\
    $\nu_\mu$       & $\left(1.2\pm0.6\right)\times10^{-11}$ & 63.7\\
    $\bar{\nu}_\mu$ & $\left(6.6\pm4.7\right)\times10^{-12}$ & 35.4\\
    $\nu_e$         & $\left(1.1\pm0.9\right)\times10^{-13}$ & 0.6\\     
    $\bar{\nu}_e$   & $\left(5\pm4\right)\times10^{-14}$ & 0.3\\
    \end{tabular}
    \end{ruledtabular}
\end{table}
While $\Phi_\nu$ is made up of 93.6\% $\nu_\mu$, 5.9\% $\bar{\nu}_\mu$, and 0.5\% 
$\nu_e,\bar{\nu}_e$~\cite{AguilarArevalo:2008yp}, $\Phi_\text{Off}$ is composed of 63.7\% $\nu_\mu$, 
35.4\% $\bar{\nu}_\mu$, and 0.9\% $\nu_e,\bar{\nu}_e$. 
The breakdown of $\Phi_\text{Off}$ by source material that the secondary beam (Fig.~\ref{fig:fig1}) 
was generated in is 55\% air, 30\% beryllium, 10\% steel, 3\% aluminum, and 2\% concrete. 
Air and beryllium provide approximately equal contributions to $\Phi_\text{Off}$ for $E_\nu$ above 500\MeV 
wth almost no contributions from the other materials.

The sensitivity to dark matter depends on the number and distribution of $\pi^0$s generated in the beam line. 
Table~\ref{tab:tab2} gives the total number of $\pi^\pm$ per POT as well as the breakdown by material 
in the beam line for both off-target and neutrino running simulated by \textsc{BooNEG4Beam}.
\begin{table}[htbp!]
\caption{\label{tab:tab2}The breakdown of the number of charged pions per POT and by material in the beam line. 
A pion was counted if it had a total kinetic energy greater than 1\MeV, was traveling in the forward direction, and 
had a transverse momentum less than 1\GeVc. 
Off-target in this table refers to the nominal beam configuration measured by the multiwires, not the average of the 
four possible scraping scenarios that is used as the off-target neutrino flux.}
\begin{ruledtabular}
      \begin{tabular}{rrr}
        & $\pi^{+}$ & $\pi^{-}$ \\\\
        \multicolumn{1}{l}{\textbf{Off-Target} $\text{meson}/\mathrm{POT}$} &  2.48 & 2.36 \\
        \multicolumn{3}{l}{Composition}\\
        Air         &  3.6\%  & 3.0\%\\
        Aluminum   &  0.2\%  & 0.2\%\\
        Beryllium   &  0.2\%  & 0.2\%\\
        Concrete    &  3.6\%  & 4.1\%\\
        Dolomite    &  0.1\%  & 0.1\%\\
        Steel       &  92.3\% & 92.4\%\\\\
        \multicolumn{1}{l}{\textbf{Neutrino Mode} $\text{meson}/\mathrm{POT}$} & 2.54 & 2.51 \\
        \multicolumn{3}{l}{Composition}\\
        Air         &  1.7\%  & 1.4\%\\
        Aluminum   &  5.3\%  & 5.2\%\\
        Beryllium   &  29.5\% & 27.6\%\\
        Concrete    &  28.0\% & 27.6\%\\
        Dolomite    &  0.1\%  & 0.2\% \\
        Steel       &  35.4\% & 38.0\%\\
      \end{tabular}
\end{ruledtabular}
\end{table}
The simulated $\pi^0$ distribution was chosen as the average of the $\pi^+$ and $\pi^-$ distributions which has been 
shown to be in good agreement with actual $\pi^0$ distributions~\cite{Singh:1972hq,Amaldi:1979zk,Jaeger:1974pk}. 
Neutrino-mode charged-pions are generated evenly in beryllium, steel, and concrete. 
The concrete surrounds the decay pipe and the steel is primarily located in the beam dump. 
The charged pions generated in the concrete and steel, if able to decay, will produce low-energy neutrinos and therefore do 
not contribute much to the on-target neutrino event rate. 
Off-target charged pions are predominantly produced in steel, which is consistent with the reduction of the neutrino flux. 
The different scraping scenarios that generate the off-target central value flux changes the number of pions produced in the steel beam dump 
by less than a percent. 

\begin{figure}[htbp!]
\subfloat[\label{fig:fig11a}\nuModeLong]{\includegraphics[width=0.25\textwidth]{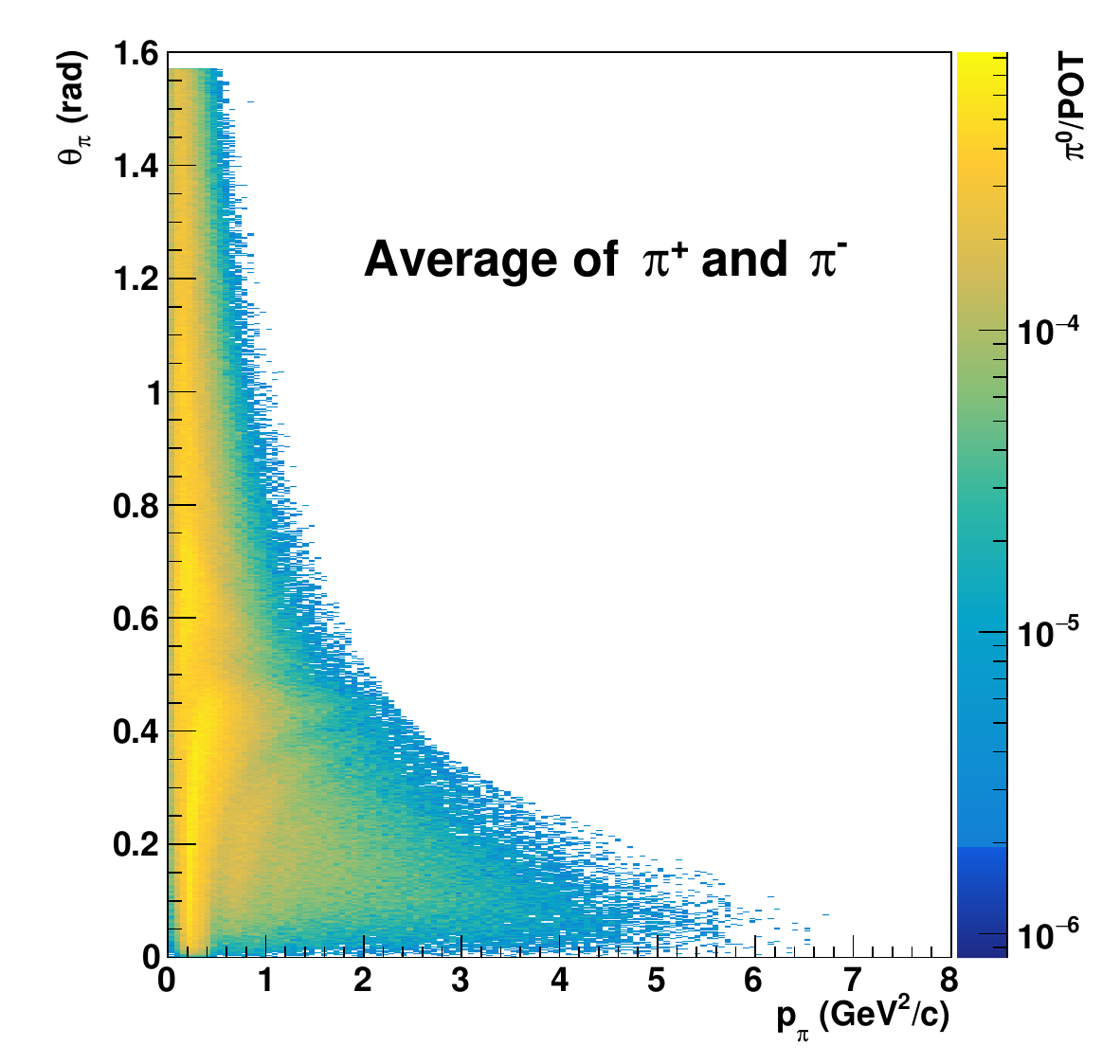}}
\subfloat[\label{fig:fig11b}off-target]{\includegraphics[width=0.25\textwidth]{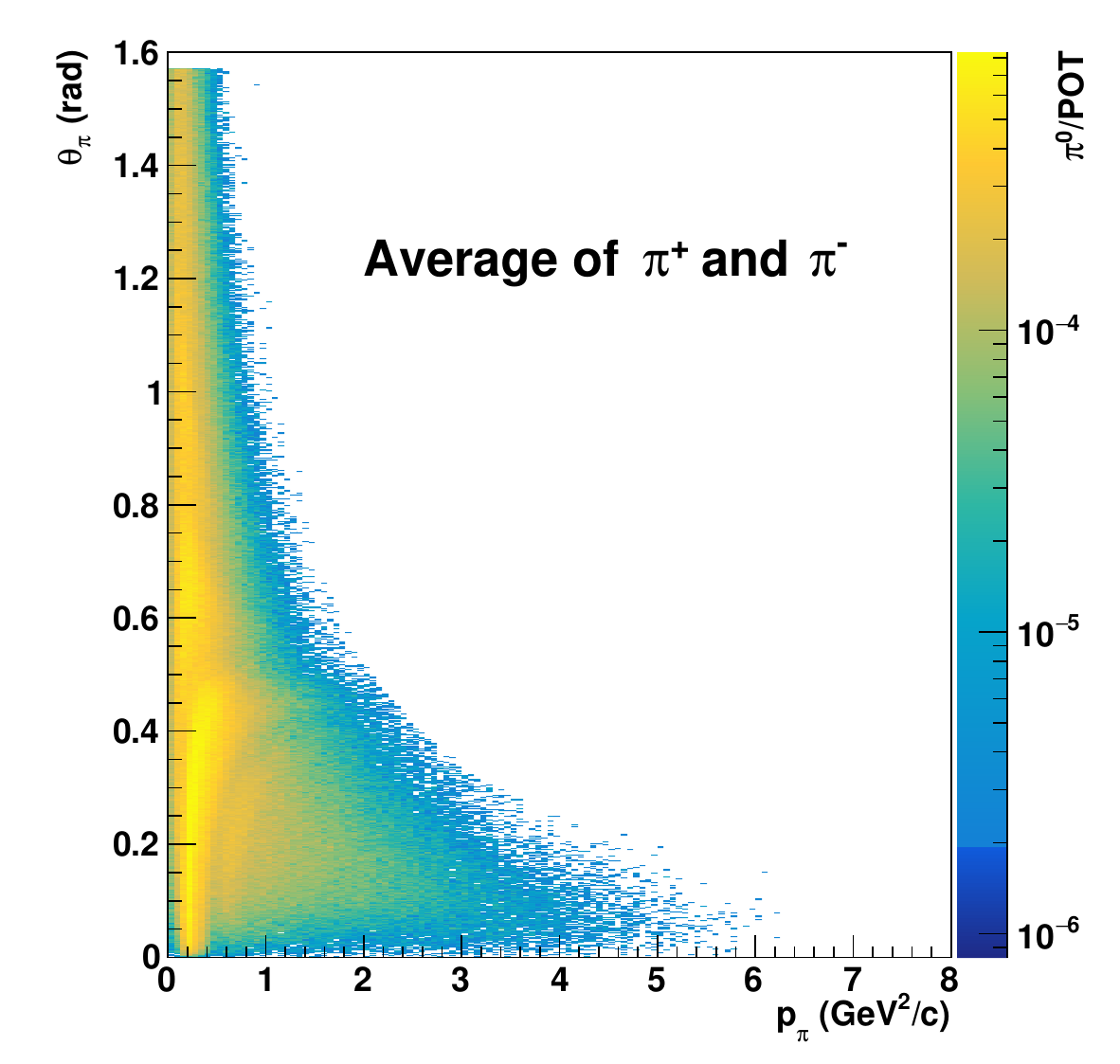}}
\caption{\label{fig:fig11}The $\theta_{\pi^0}$ vs $p_{\pi^0}$ distributions from \textsc{BooNEG4Beam} used 
for generating dark matter candidate events. 
The color scale gives the number of pions per delivered POT in each bin.}
\end{figure}

Taking the average of the charged-pion distributions to generate the $\pi^0$ distribution, the off-target $\pi^0$ distribution 
will generate a greater dark matter flux than on-target because more of the pions are generated at the beam dump 
transversely closer to the center of the beam spot. 
Figure~\ref{fig:fig11} shows the angle vs total pion momentum for the $\pi^0$ distribution used as 
input to the \DMS simulations, discussed in Sec.~\ref{sec:modelDep}. 
The total integral is dominated by low-momentum pions, where pions were simulated down to a total kinetic energy of 1\MeV. 

For the $\eta$-meson distribution the $\pi^0$ distribution was reweighted by setting the total momentum of the $\eta$ meson to be
\[
p_\eta = \sqrt{E_{\pi^0}^2 - m_\eta^2},
\]
where $E_{\pi^0}$ is the total energy of the $\pi^0$ being reweighted, and $m_\eta$ is the mass of the $\eta$-meson. 
Only $\pi^0$ events that satisfy $E_{\pi^0} > m_\eta$ were used in the reweighting scheme. 
The momentum vector $\mathbf{p}$ for the $\eta$ meson is then calculated by
\[
\mathbf{p}_{\eta} = \mathbf{p}_{\pi^0}\frac{p_\eta}{p_{\pi^0}}.
\]
A systematic test was preformed to generate the $\eta$ meson distribution by doing the same procedure above but starting
with the predicted off-target kaon distribution instead of the $\pi$ distribution.
An independent simulation using \textsc{Pythia}~\cite{Sjostrand:2006za} predicted the $\eta$ distribution to closely match the distribution obtained 
by reweighting of the kaon distribution.
The final confidence level limits, discussed in Sec.~\ref{sec:modelDep}, showed no change in the predicted sensitivity for
a slice of the dark matter parameter space. The predicted $\eta$ distribution used for the final analysis was the one 
that used the $\pi$ distribution, because there are smaller uncertainties on the $\pi$ production.
 
A particle list of $\pi^0$ and $\eta$ mesons with their 4-momentum and 4-position information was passed to the dark 
matter simulation (see Sec.~\ref{sec:modelDep}) as input for neutral meson production of dark matter.

\subsection{\label{sec:rwmTiming}Bunch time}
As the beam travels down the beam line the protons induce image charge on the vacuum pipe. 
A resistive wall current monitor (RWM) right in front of the beryllium target uses the image charge to measure the longitudinal 
bunch shape and the time the bunch hits the target~\cite{Fellenz:1998qy,Backfish:2013/05/14aya}.
The RWM design is based on the RWMs that were installed in the Fermilab Main Injector.
The intensity of the individual 2\ns bunches are measured to 1\% and the timing is known to less than 1\ns.
The RWM signals, one for each bunch, are saved for each data acquisition (DAQ) window (described in the next section). 

\begin{figure}[htpb!]
\includegraphics[width=0.5\textwidth]{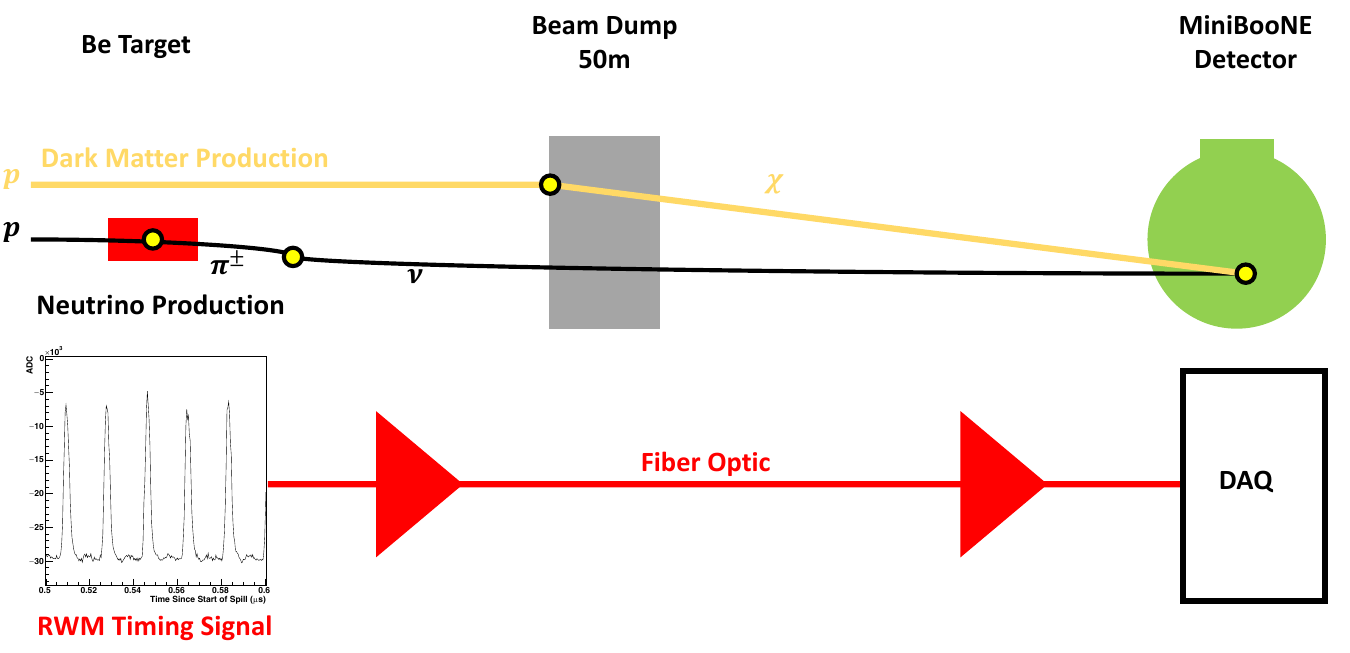}
\caption{\label{fig:fig12}Illustration showing how the RWM time signature is passed to the detector. 
Production of neutrinos and dark matter particles are also shown for comparison. 
Heavy dark matter will arrive later than the neutrinos.}
\end{figure}

\begin{figure}[htbp]
  \includegraphics[width=0.5\textwidth]{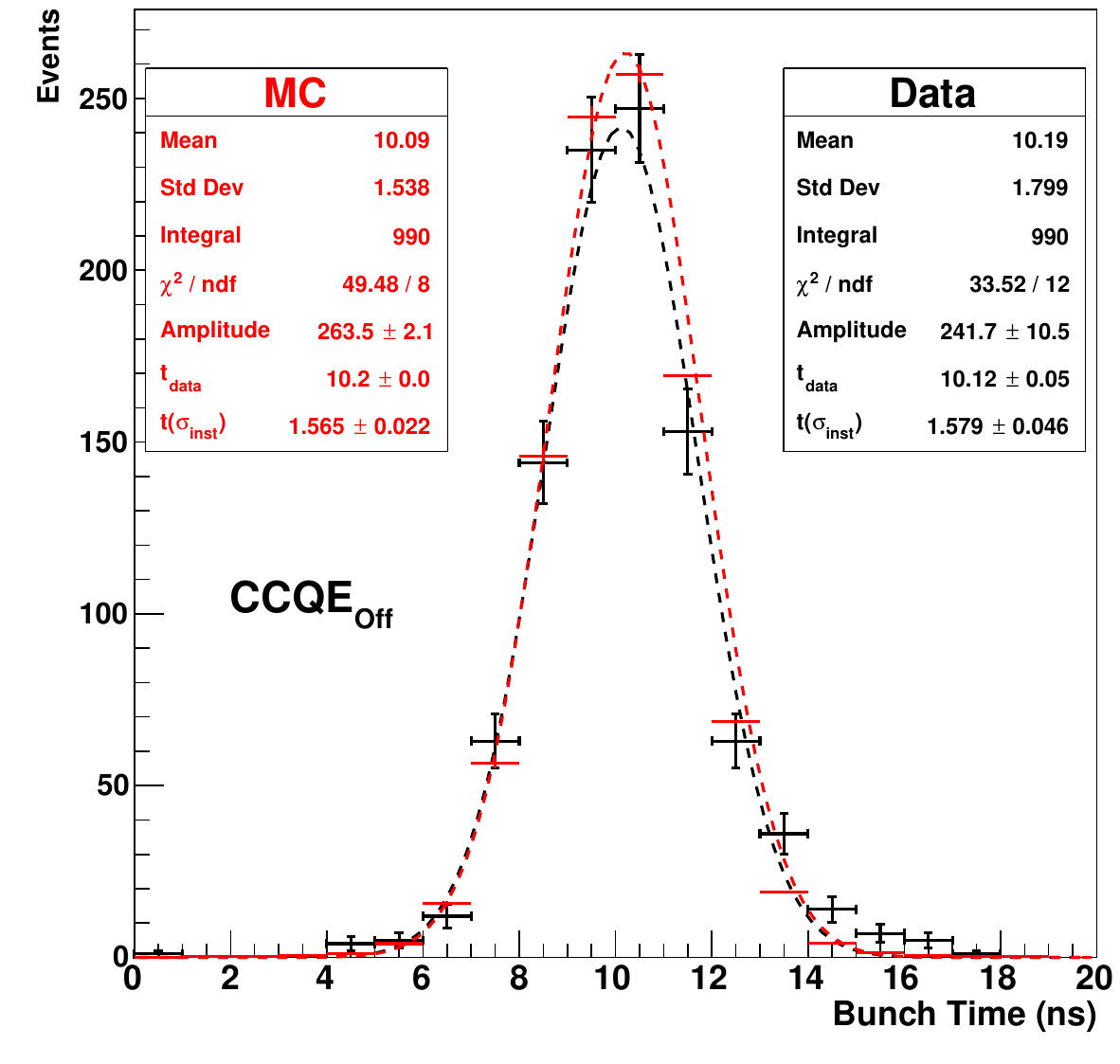}
  \caption{\label{fig:fig13}Comparison of simulated and measured \CCQE bunch times after applying 
  $\delta t\left(\sigma_\text{inst}\right)$ and $\delta t_\text{data}$ calibrations (see text). Only statistical errors are shown.}
\end{figure}

The RWM signal is sent to the \MB DAQ by an optical fiber, as shown schematically in Fig.~\ref{fig:fig12}.
For each reconstructed event a time is calculated to the first RWM bunch that passes threshold, as shown 
in Fig.~\ref{fig:fig4}. 
The bunch time is the remainder of the time of the event subtracted by the time of the RWM divided by 18.9\ns, 
see Fig.~\ref{fig:fig13}. 
The measured bunch time is a time-of-flight measurement, where two regions are defined, in-time and out-of-time. 
The in-time region is between 5.6 and 14.5\ns determined from the off-target \CCQE data mean and standard deviation. 

\begin{figure}[htbp!]
\includegraphics[width=0.5\textwidth]{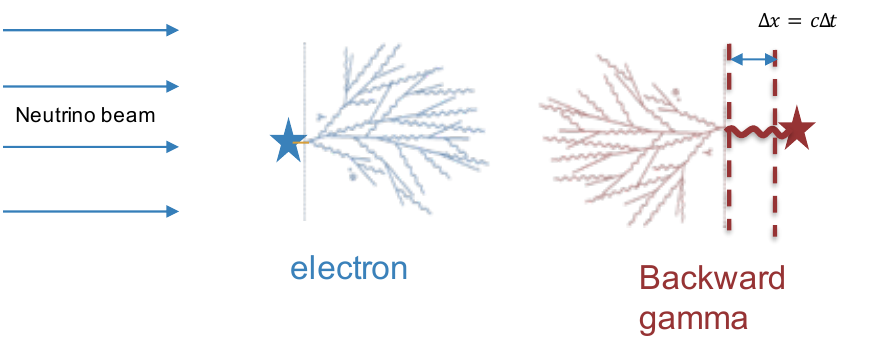}
\caption{\label{fig:fig14}Illustration of the timing difference between an electron event and a backward-going photon.}
\end{figure}

Cherenkov light has a timing resolution of $\sim1.5\ns$, while the timing resolution of scintillation light is $\sim4.2\ns$ 
from the lifetime of the scintillation light. 
The bunch times of photon events has the same timing resolution as that of muon and electron events but are shifted 
later in time because the photons travel some distance before converting into an electromagnetic shower in the detector, 
as illustrated in Fig.~\ref{fig:fig14}. 

\begin{figure}[htbp]
  \includegraphics[width=0.5\textwidth]{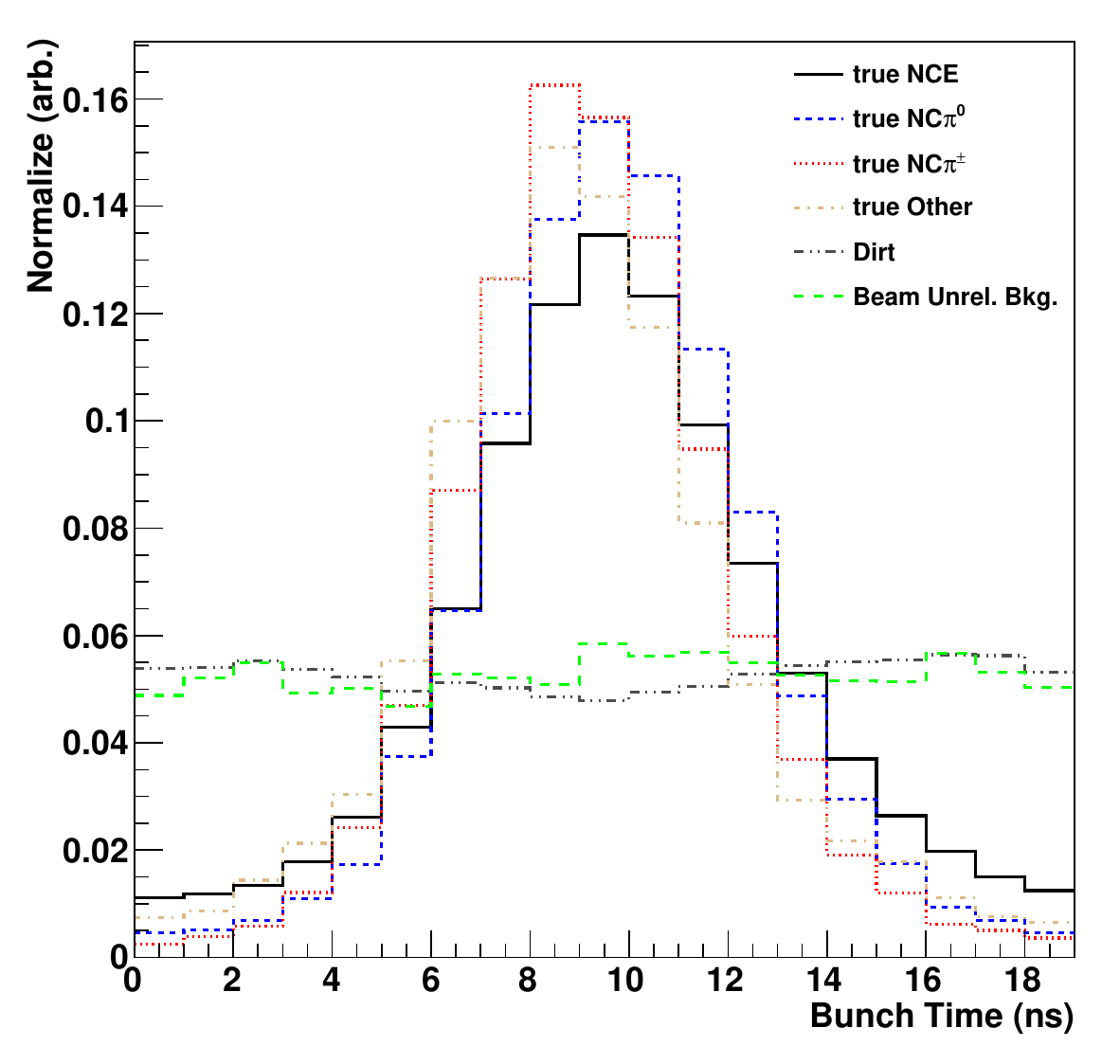}
  \caption{\label{fig:fig15}Comparison of the bunch-time shape for different event types, that pass \NCE 
  selection cuts, as predicted by the detector simulation.}
\end{figure}

The beam-unrelated backgrounds and beam-related events that happen outside the detector (\dirtBRB) have 
flat distributions in bunch time, as shown in Fig.~\ref{fig:fig15}. 
This allows for an analysis cut to remove more background events or to look for a bump in the out-of-time region for new physics. 
The timing information could also be used as an extra particle identification parameter, because different event types, or 
even final-state particles, have different shapes in bunch time for the same selection cuts, as shown in Fig.~\ref{fig:fig15}.

\subsubsection{\label{sec:rwmTiming_MC}Simulating bunch time}
The simulated bunch time $T_\text{bunch}$ was calculated by,
\[
T_\text{bunch} = \delta t_\text{reco} - \delta t_Z + \delta t_\text{dcy}  - \delta t\left(\sigma_{RF}\right) - 
\delta t\left(\sigma_\text{inst}\right) - \delta t_\text{data},
\]
where $\delta t_\text{reco}$ is the difference between true and reconstructed time, $\delta t_Z$ is the time it takes 
light to travel from the $z=0$ plane to the plane the event occurred in, and $\delta t_\text{dcy}$ is the difference between 
the time it takes light to get from the target to where the neutrino occurred and the simulated decay chain time. 
$\delta t\left(\sigma_{RF}\right)$ is a number based on the time jitter of the radio-frequency bunch structure, 
measured to be 1.15\ns. 
$\delta t\left(\sigma_\text{inst}\right)$ is the jitter from \MB instrumentation and $\delta t_\text{data}$ is the mean bunch-time 
difference between simulation and data. 
Both $\delta t\left(\sigma_\text{inst}\right)$ and $\delta t_\text{data}$ were tuned to off-target \CCQE data, as shown in 
Fig.~\ref{fig:fig13}.

If \DMS has a mass approximately greater than 25\MeVcc, it could reach the detector in the out-of-time region. 
This would distort the bunch time distribution. 
The bunch time is used in this analysis as an extra constraint on the possible dark matter parameters.


\section{\label{sec:detector}\MB detector}
The \MB detector, described in Ref.~\cite{AguilarArevalo:2008qa}, is a Cherenkov and scintillation tracking detector 
designed to search for $\nu_e$ and $\bar{\nu}_e$ appearance oscillations at short baseline~\cite{Aguilar-Arevalo:2013pmq}.
It is located 541\m downstream from the center of the BNB neutrino target. 
As described above, the majority of dark matter production is expected to occur at the 50-m absorber whose front face is 
491\m from the detector center. 
The proton beam is aligned 1.9\m below the center of the detector during normal neutrino running. 

The detector is a 12.2\m diameter spherical tank filled with 818 tons of Markol 7 light mineral oil (C$_n$H$_{2n+2}$ 
where $n \approx 20$). 
No additives were introduced in the mineral oil, but there remain small levels of fluorescent contaminants. 
There is a spherical optical separation with a radius of 5.476\m centered within the main volume. 
The outer ``veto'' region contains the same mineral oil as the inner ``tank'' region despite being optically separated. 

The index of refraction of the oil was measured to be 1.47, yielding a Cherenkov light threshold for particles with $\beta > 0.68$. 
For protons (electrons) this is approximately 280\MeV (150\keV). 
The impurity fluors contribute enough scintillation light to push the proton detection threshold well below this. 

The inner region is viewed by 1280 inward-facing 8-inch photomultiplier tubes (PMTs).  
These PMTs are mounted on the inner surface of the optical barrier and provide 11.3\% photocathode coverage. 
The outer region is viewed by 240 PMTs arranged in pairs around the outside of the optical barrier. 
These outer-region PMTs are of the same type as the inner region. 

The light signal read out by the PMTs is sent to custom-built electronics where the signal is amplified, discriminated, 
and then digitized. 
The electronics (``QT'' boards~\cite{Athanassopoulos:1996ds}) both integrate the total charge and extract the start 
time of the digitized pulse.
Threshold was equivalent to about 0.1 photoelectrons. 
All the hits from all the PMTs are accumulated into buffers to await a trigger decision from the logic. 
The multiplicity of PMT hits and external signals are used to create various triggers for physics and calibrations. 

When a trigger occurs, 19.2\,$\mu$s of PMT hits are extracted from the QT boards. 
The physics trigger was a Fermilab accelerator signal that signals when protons are being delivered to the BNB area. 
The 1.6\,$\mu$s beam spill is placed 5\,$\mu$s after the start of data acquisition. 
Therefore, the intrinsic cosmic-ray background activity is adequately measured.  
The remaining 12\,$\mu$s of time measures the neutrino-induced muon decays which have a lifetime of 2.2\,$\mu$s. 

\subsection{\label{subsec:detector_sim}Detector simulation}
The detector simulation was split up into neutrino interaction and detector response. 
The neutrino interaction simulation used a modified version of the \NUANCE \textsc{V3} neutrino event generator for 
simulating neutrino interactions in CH$_2$~\cite{CASPER2002161}. 
Descriptions of the relevant \NUANCE models and uncertainties were given in 
Refs.~\cite{AguilarArevalo:2010cx} and~\cite{AguilarArevalo:2010zc} for \NCE and \CCQE respectively and in 
Ref.~\cite{AguilarArevalo:2009ww} for neutral-current single $\pi^0$ production (\NCPi). 
In summary, the relativistic Fermi gas model of Smith and Moniz is used to describe both \NCE and \CCQE events, 
while the Rein and Sehgal models~\cite{Rein:1980wg,Rein:1982pf} are used to predict \NCPi. 
Pion absorption and charge exchange are included in generating the final-state particles.
The axial form factor is assumed to be of dipole form with an axial mass $M_a$ and a Pauli-blocking parameter $\kappa$ 
is introduced as an extra degree of freedom to model low 4-momentum transfer $Q^2$ events in 
\MB correctly~\cite{AguilarArevalo:2010zc}. 

\MB used $M_A^\text{eff} = 1.23 \pm 0.20\GeVcc$ and $\kappa = 1.019\pm0.011$ for the simulations generated for
 \nuModeLong publications. 
In Ref.~\cite{AguilarArevalo:2010zc} $M_A^\text{eff}$ and $\kappa$ were measured to be $1.35\pm0.17\GeVcc$ and 
$1.007\pm0.012$, respectively, with an extra 1.08 normalization factor to match simulations with data. 
For this analysis all detector and dirt simulated events were reweighted to these updated measured values, while only 
true \CCQE events include the normalization factor.

The detector response is modeled with a \textsc{Geant3}~\cite{Brun:1987ma} simulation described in 
Ref.~\cite{AguilarArevalo:2008qa}.

\subsubsection{\label{subsubsec:nce_def}Definition of true interactions}
The dark matter simulation (\BdNMC) that is used (Sec.~\ref{sec:modelDep}) does not include a nuclear model or final-state 
interactions. In order to connect the \NUANCE and detector simulations to \BdNMC, ``true'' neutrino
interactions are defined by the output of a neutrino interaction before any final state interactions or nuclear model are considered. This
makes the definitions used by \BdNMC and \NUANCE
the same.

It should be noted that the weighting scheme to produce a predicted dark matter spectrum coming from the
detector simulations is discussed in Sec.~\ref{sec:modelDep}. The procedure applies the nuclear model and the model for final-state 
interactions that are in \NUANCE to \BdNMC to correctly determine the reconstructed dark matter distribution.


\begin{table}[htbp!]
  \caption{\label{tab:tab3}Selection cuts for the various channels in this analysis}
  \begin{ruledtabular}
    \begin{tabular}{rl}
      Cut \# & \multicolumn{1}{c}{Description} \\
      \hline\\
     
     \textbf{\CCQE}\\
      1                   & \# subevents = 2                                               \\
      2                   & 1st sub, \# tank $>$ 200 and       \\
                          & all subevents, \# veto hits $<$ 6 \\
      3                   & 1st sub, reconstructed vertex radius $< 500\cm$                \\
      4                   & 1st sub, event time window $4.4 < T\left(\us\right) < 6.4$     \\
      5                   & 1st sub, $\mu/e\text{ log-likelihood ratio} > 0$               \\
      6                   & 1st sub,  kinetic energy $T > 200\MeV$                         \\
      7                   & $\mu$-e vertex distance $> 100\cm$ and                         \\
                          & $> \left(500 T_\mu\left(\GeV\right) - 100\right)\cm$           \\\\
     \textbf{\NCE}\\
      1                   & \# subevents = 1 \\
      2                   & \# tank hits $>$ 12 and \# veto hits $<$ 6 \\
      3                   & Reconstructed vertex radius $< 500\cm$ \\
      4                   & event time window $4.4 < T \left(\us\right) < 6.4$ \\
      5                   & $p$/$e$ time log-likelihood ratio $< 0.42$ \\
      6                   & kinetic energy $35 < T\left(\MeV\right) < 650$ \\
      7                   & $<$ 60 hits 10\us before event trigger\\\\

      \textbf{\NCPi}\\
      1                   & \# subevents = 1                                               \\
      2                   & \# tank hits  $>$ 200  and \# veto hits  $<$ 6                 \\
      3                   & event time window $4 < T \left(\us\right) < 7$                 \\
      4                   & Reconstructed vertex radius $< 500\cm$ (e fit)                 \\
      5                   & $\mu/e\text{ log-likelihood ratio} > 0.05$                     \\
      6                   & $e/\pi^0\text{ log-likelihood ratio} < 0$                      \\
      7                   & $80 < m_{\gamma\gamma}\left(\MeVcc\right) < 200$               \\\\

     \textbf{\NCEE}\\
      1                   & \# subevents = 1\\
      2                   & \# tank hits $>$ 20 and \# veto hits $\le$ 2 \\
      3                   & event time window $4.4 < T \left(\us\right) < 6.4$ \\
      4                   & Reconstructed vertex radius  $< 500\cm$ \\
      5                   & visible energy $75 \le \EvisE \left(\MeV\right) \le 850$ \\
      6                   & reconstructed angle $\costhetae \ge 0.9$\\
      7                   & $\mu/e\text{ log-likelihood ratio}$: See text\\
      8                   & $e \text{ time log-likelihood} \le 3.6$ \\
      9                   & Scintillation / Cherenkov Ratio $\le$ 0.55 \\
      10                  & Distance to wall $\ge$ 210\cm \\\\
                          & For events with \# tank hits $>$ 200 \\
      11                  & $e/\pi^0\text{ log-likelihood ratio} > -6.25\times10^{-3}$\\
      12                  & $m_{\gamma\gamma}\le 80\MeVcc$ \\
    \end{tabular}
  \end{ruledtabular}
\end{table}

\section{\label{sec:distributions}Event Distributions}
This analysis consists of four different selection cuts: \CCQE, \NCE, \NCPi, and neutral-current neutrino-electron 
elastic scattering (\NCEE).
Because of final-state interactions the events that pass these selection cuts are \CCQE-like, \NCE-like, \NCPi-like, and \NCEE-like events. 
For simplicity, for the rest of this paper we will leave off the ``-like'' when referring to the events that passed the cuts.
\CCQE candidate events, defined by seeing a primary muon followed by the decay electron, are used to determine the neutrino event rate.
\NCE, \NCPi, and \NCEE are considered signal channels.
Table~\ref{tab:tab3} gives selection criteria for each selection cut.

\begin{figure}[htbp!]
\includegraphics[width=0.5\textwidth]{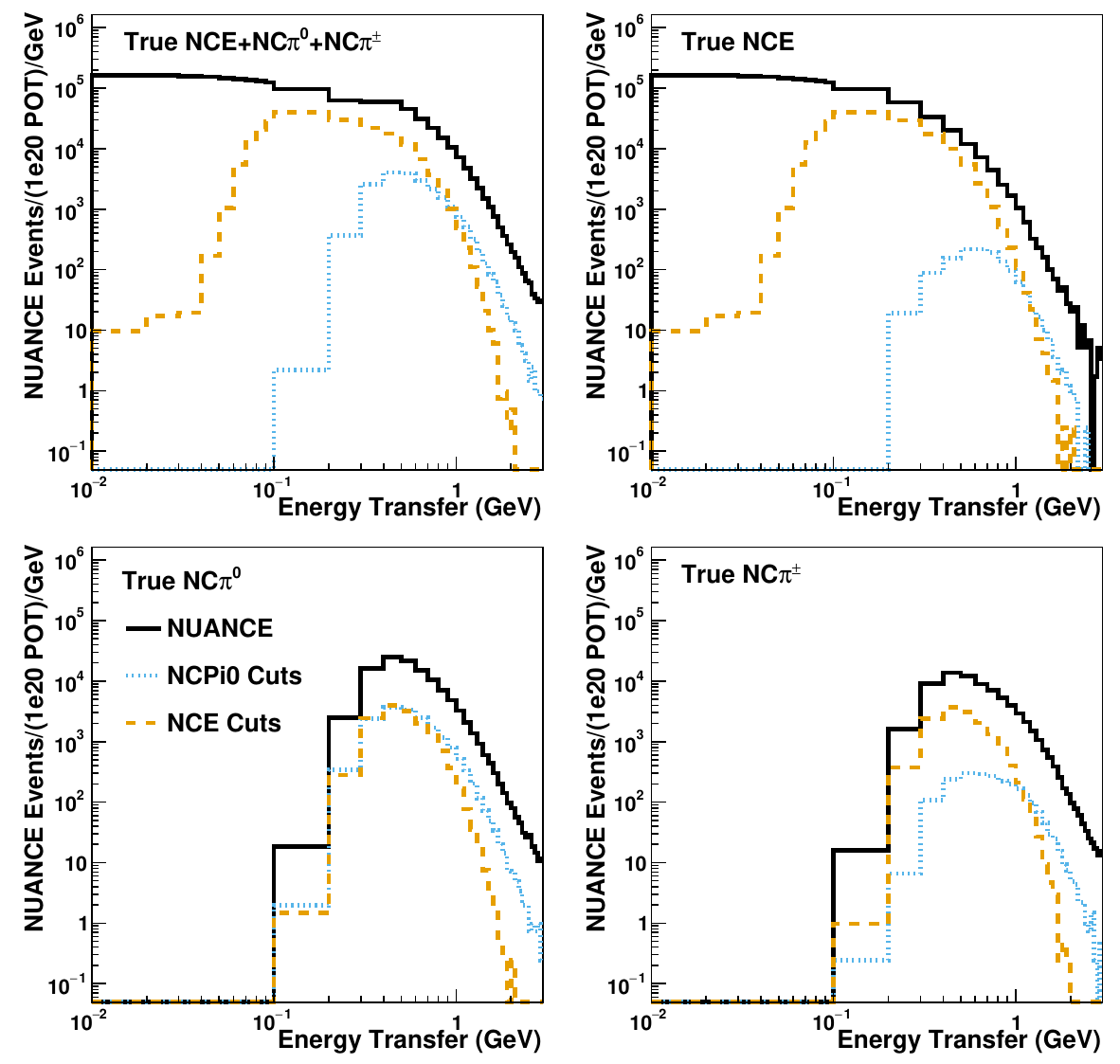}
\caption{\label{fig:fig16}The expected number of events before and after \NCE and \NCPi cuts for true \NCE, true 
\NCPi and true \NCPip interactions. \NCE cuts are just as efficient at detecting true \NCPi events at low energy transfer 
because of the pion absorption in the nucleus.}
\end{figure}
\begin{table}[htpb!]
\caption{\label{tab:tab4}The efficiency for \NCE and \NCPi selection cuts for different true signal channels. 
Only true events that interact in the fiducial volume $r < 500$\cm are considered. 
For true \NCE interactions the efficiency given in the parentheses is for energy transfer $>90\MeV$.}
\begin{ruledtabular}
\begin{tabular}{rll}
\multirow{2}{*}{True Interaction} & \multicolumn{2}{c}{Selection Cut Efficiency (\%)}\\
& \multicolumn{1}{l}{\NCE} & \multicolumn{1}{l}{\NCPi}\\
\NCE & 11 (37) & $<$0.1 (0.4)\\
\NCPi & 13 & 15 \\
\NCPip & 20 & 3 \\
Total & 12 (30) & 1 (4) \\
\end{tabular}
\end{ruledtabular}
\end{table}

\NCE cuts follow from the \MB antineutrino-\NCE analysis~\cite{Aguilar-Arevalo:2013nkf} with the addition of \NCE{7}, a 
previous trigger activity cut.
A subevent is defined as a group of hits where no hits are separated by more than 10\,ns and the group has no less 
than 10 hits. 
Only a single nucleon is expected for \NCE interactions, which is why the \NCE{1} cut allows only one subevent within 
the 19.2\us DAQ window.
\NCE{2} makes sure that the subevent falls within the beam window inside the DAQ window.
\NCE{3} requires a minimal number of tank hits needed for reconstruction and a maximum number of veto hits for beam-unrelated background 
rejection.
\NCE{4} sets the fiducial volume, and \NCE{5} separates scintillation-like events from Cherenkov-like events based on 
the response time of each process.
\NCE{6} selects the kinetic energy parameter space for the analysis.
\NCE{7} is used to further remove beam-unrelated backgrounds by looking at events that happen in the detector before the beam trigger turned on. 
The cut is set to reject all events that have a trigger of greater than or equal to 60 hits in the detector within 10\,$\mathrm{\mu s}$ 
before the event trigger. 
The efficiency of \NCE{7} for beam-related events that passed the previous cuts is $\left(95.3\pm0.2\right)\%$, while beam-unrelated backgrounds 
are reduced by 42.5\%~\cite{Thornton:2017etu}.

\NCE selection cuts are 30\% efficient at detecting true \NCE, \NCPi and neutral-current single charged pion (\NCPip) events 
with an energy transfer greater than or equal to 90\MeV and that interact in the fiducial volume. 
The effects below 90\MeV were discussed in the previous section.
\NCE selection cuts result in 95\% pure true \NCE, \NCPi, and \NCPip events. Table~\ref{tab:tab4} gives the breakdown of 
efficiency for the different true channels, and Fig.~\ref{fig:fig16} shows the efficiency as a function of energy transfer. 

True \NCEE events are very forward. 
A $\costhetae > 0.9$ cut was used to have a control region to estimate the background distribution in the signal region, 
defined by $\costhetae > 0.99$.
The \NCEE selection cuts have a stricter number of veto hits (\NCEE{2}) than \NCE along with a distance to the wall cut 
(\NCEE{10}) to remove \dirtBRB events. 
The selections \NCEE{8} and \NCEE{9} are used to reduce the \NCE background. 
The selection \NCEE{7} rejects muon background and uses the same values as that for the oscillation 
analysis~\cite{Patterson:2009ki,PattersonThesis}. 
The selections \NCEE{11} and \NCEE{12} are used to remove high-multiplicity events with a $\pi^0$.
Events with less than 200 tank hits automatically pass \NCEE{11} and \NCEE{12} for the cuts are only applied to high 
multiplicity events.

The selected \NCEE distribution is beam-unrelated background free because of \NCEE{5}, which sets the minimal reconstructed visible energy 
\EvisE above the end point of the electron from muon decay. 
The high-\EvisE cut in \NCEE{5} was tuned to maximize efficiency times purity of the \NCEE sample in the signal region.
Lowering the \EvisE will allow more of the predicted dark matter, but the increase in the beam-unrelated backgrounds decreases the sensitivity. 
The \NCEE selection cuts are 15\% efficient and 63\% pure in the signal region for true \NCEE events that interact inside 
the fiducial volume.

About 40\% of the \NCEE candidate events, $\costhetae > 0.9$, also pass the neutrino oscillation selection cuts as 
employed in previous analyses~\cite{Patterson:2009ki,Aguilar-Arevalo:2013pmq}. 
The majority of the events that pass \NCEE but not oscillation selection cuts come from the lower number of tank hits and 
\EvisE cuts, along with having no $\pi^0$ cuts for events with less than 200 tank hits. 
Applying the neutrino oscillation selection cuts to the off-target data is discussed in Sec.~\ref{sec:osc_events} and Sec.~\ref{sec:osc}.

The cuts for \CCQE candidate events are similar to the cuts from Ref.~\cite{AguilarArevalo:2010zc} with the addition of 
the 200-tank hit cut on the first subevent.
The cuts for \NCPi candidate events are the same as the cuts from Ref.~\cite{AguilarArevalo:2009ww}, except for a 
wider event timing cut. 
\NCPi selection cuts are 4\% efficient and 86\% pure for true \NCE, \NCPip, and \NCPi events. 
See Table~\ref{tab:tab4} and Fig.~\ref{fig:fig16} for the breakdown by true interaction channel and as a function of 
energy transfer.

The subscripts $\nu$, $\bar{\nu}$, or ``off'' will be added to the distribution label when specifically mentioning events after 
passing cuts from neutrino, antineutrino, or off-target modes respectively.

\subsection{\label{subsec:backgrounds}Backgrounds}
There exist two categories of backgrounds: beam-related and beam-unrelated.
The beam-unrelated backgrounds are measured by a 2\,Hz (10-15 Hz) random trigger for on-target (off-target) running, and scaled by the ratio of 
number of beam triggers with POT delivered to the number of random triggers.
Knowing that beam-unrelated backgrounds were going to be more significant , then random trigger data-taking rate was increased for off-target running. 
Beam-related backgrounds can be further split into events that occur in the detector and \dirtBRB events (see Sec.~\ref{sec:rwmTiming}).
For this analysis, all neutrino interactions were considered background.

\begin{figure*}[htbp!]
\includegraphics[width=\textwidth]{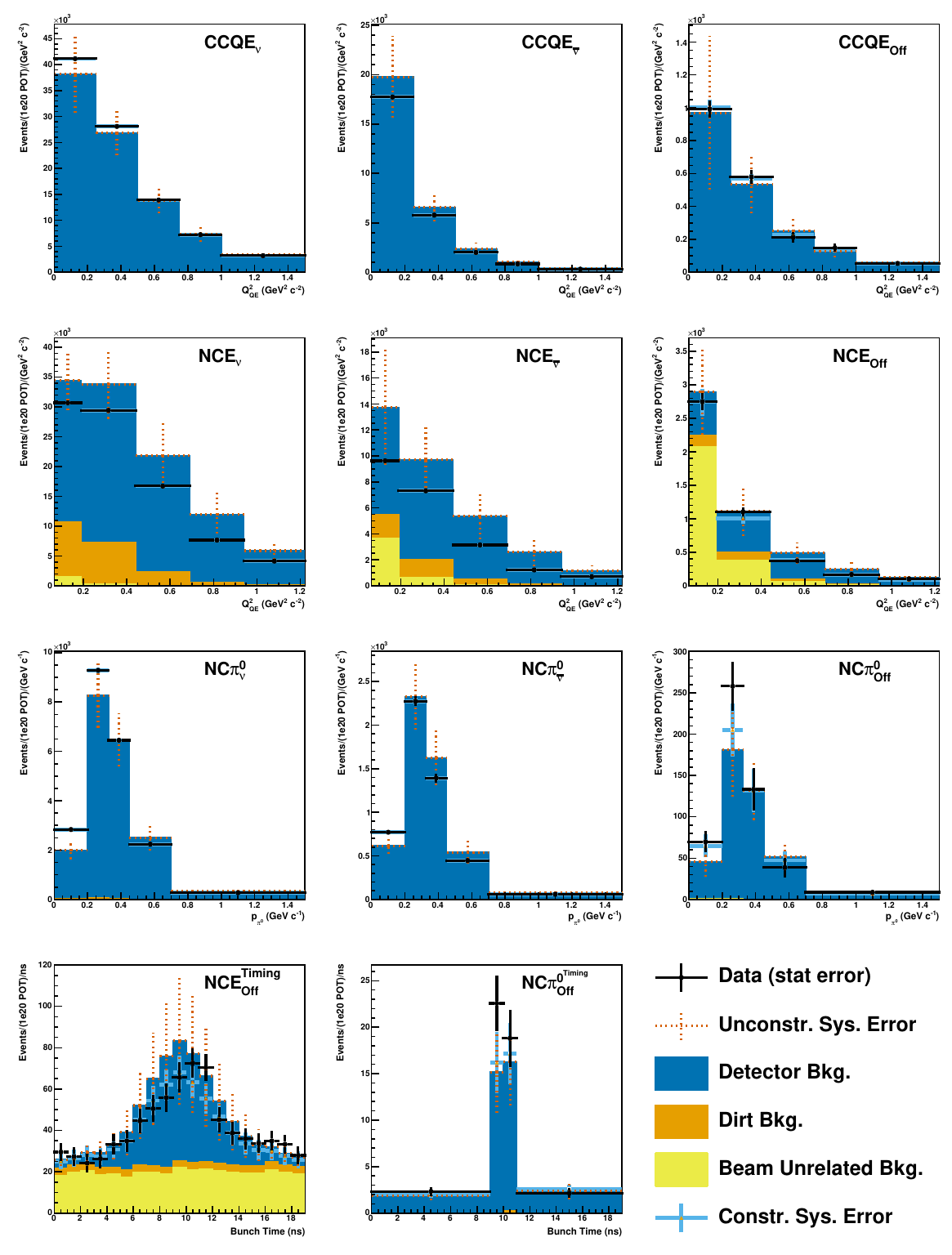}
  \caption{\label{fig:fig17}\CCQE, \NCE, \NCPi distributions for neutrino, antineutrino and off-target modes. 
  The points are data with statistical errors. The triangles with light lines are constrained predictions with constrained systematic errors. 
  The dotted lines are the total predictions with unconstrained systematic errors.
  The backgrounds that add up to give the total predictions are given by the stacked histogram.}
\end{figure*}

\begin{figure}[htbp!]
\includegraphics[width=0.5\textwidth]{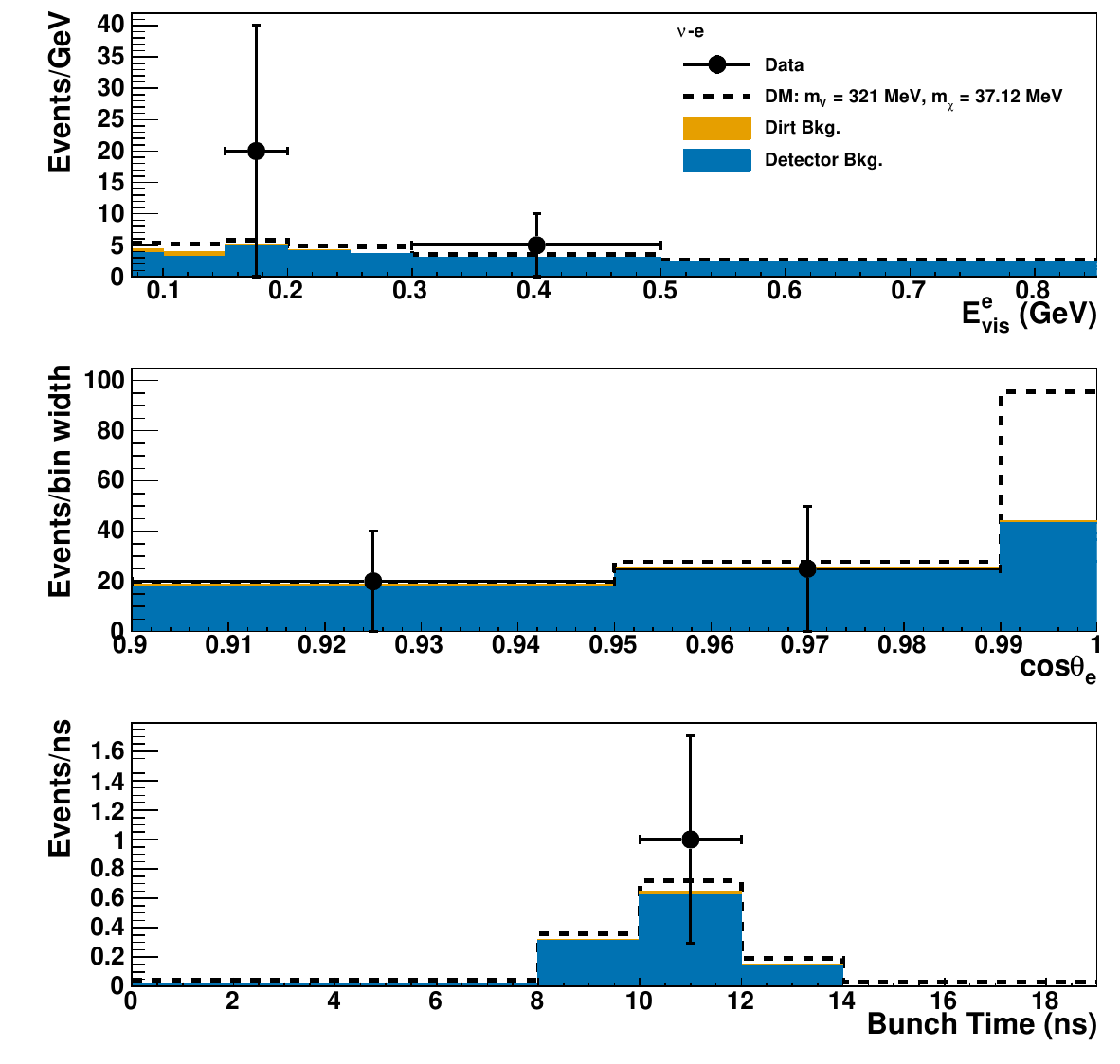}
  \caption{\label{fig:fig18}The (top) visible electron energy \EvisE, (middle) electron angle \costhetae, and (bottom) 
  bunch time distributions that pass \NCEE cuts for off-target mode. The prediction was scaled to match the number of data events 
  for $0.9 \le \costhetae < 0.99$. 
  An example dark matter prediction is given (dashed line) to illustrate how forward the resulting electron is expected to be.}
\end{figure}

Beam-unrelated backgrounds were overlaid on top of simulated beam events to correctly simulate the rejection of beam events that have beam-unrelated backgrounds in 
the same DAQ window.
The rate of events passing the one subevent and number of veto hits less than six cuts from the random trigger increased 
by 3.8\% from neutrino mode to off-target mode. 
Figure~\ref{fig:fig17} shows the number of events as a function of the dependent variables for \CCQE, 
\NCE, \NCPi broken down by predicted background. Also shown are the timing distributions for \NCEOff and \NCPiOff.
The systematics shown in Fig.~\ref{fig:fig17} are the total systematic uncertainties before constraints 
are applied (see Sec.~\ref{subsec:systematics}).
Looking at the three \NCE distributions the relative percentage of beam-unrelated backgrounds increases as the neutrino interaction rate decreases.
The resulting \NCEE distributions are shown in Fig.~\ref{fig:fig18}.

\subsection{\label{subsec:systematics}Systematic uncertainties}
The study of systematic uncertainties considered the correlations between the \NCE, \NCPi, and \CCQE distributions for neutrino, 
antineutrino, and off-target modes, as well as the timing distributions for \NCEOff and \NCPiOff  denoted by \NCEOffTiming and 
\NCPiOffTiming respectively.
Table~\ref{tab:tab5} gives the breakdown of the systematics on the total background prediction for all 
distributions considering bin-to-bin correlations and no constraints.
\begin{table}[hbtp!]
\caption{\label{tab:tab5}The total unconstrained error broken down 
by source and distribution. The total constrained error for \NCEOff is 6.4\% and 11.0\% for \NCPiOff.}
\begin{ruledtabular}
\begin{tabular}{lrrrrr}
\multirow{1}{*}{Distribution} & \multicolumn{5}{c}{Source unconstrained total uncertainty (\%)}\\
& \multicolumn{1}{c}{\multirow{2}{*}{$\nu$ flux}} & \multicolumn{1}{c}{cross} & \multicolumn{1}{c}{detector} & \multicolumn{1}{c}{total} 
& \multicolumn{1}{c}{\multirow{2}{*}{statistical}}\\
&  & \multicolumn{1}{c}{section} & \multicolumn{1}{c}{model} & \multicolumn{1}{c}{systematic} 
& \\
\multicolumn{6}{l}{\textbf{Neutrino Mode}}\\
\CCQENu   &  5.9 & 16.2 &  3.3  & 17.6 & 0.3\\
\NCENu      &  5.5 & 12.7 & 13.6 & 19.5 & 0.3\\
\NCPiNu     & 7.7  & 10.5  & 10.2 & 16.5 & 0.7\\\\
\multicolumn{6}{l}{\textbf{Anti-neutrino Mode}}\\
\CCQENuBar & 5.6 & 18.4 & 9.3 & 21.4 & 0.3 \\
\NCENuBar    & 4.7 & 16.0 & 19.7 & 27.8 & 0.4 \\
\NCPiNuBar   & 7.0 & 7.9 & 14.5 & 17.9 & 1 \\\\
\multicolumn{6}{l}{\textbf{Off-Target}}\\
\CCQEOff & 32.8 & 17.9 &  3.0 & 37.5 & 3.2\\
\NCEOff    & 25.9 &  7.7 &  7.8 & 28.2 & 2.6 \\
\NCPiOff   & 26.7 & 10.0 & 10.3 & 30.3 & 9 \\
\end{tabular}
\end{ruledtabular}
\end{table}
The total systematic uncertainty is the quadratic sum of the three categories given plus the uncertainties on the previous trigger 
activity cut and random trigger scaling.
For \CCQE and \NCPi the uncertainty on the previous trigger activity is zero and practically zero for the random trigger scaling. 
The reduction in the cross section and detector model uncertainties in \NCEOff compared to \NCENu comes from the increased 
percentage of beam-unrelated backgrounds. 
\NCE has a lower total uncertainty in cross sections compared to \CCQE because \NCE (\CCQE) is most uncertain at higher (lower) 
reconstructed 4-momentum transfer using the quasielastic assumption $Q_{QE}^{2}$ [Eqs.~(\ref{eq:eq3}) and~(\ref{eq:eq4}) 
for definitions] where there are less (more) events.

The total constrained uncertainty is calculated by considering that all nonsignal bins constrain the signal bins. 
For \NCEOff and \NCPiOff the total constrained uncertainties are 6.5\% and 11.0\% respectively. 
Statistical uncertainties are included in the total constrained calculation.  

The shape-only uncertainty is 6.8\% for \NCPiOffTiming and 2.3\% for \NCEOffTiming, and is dominated by uncertainties in the 
detector model. 
The uncertainty in the instrumentation of the RWM and calibration of the simulation is small compared to the uncertainty from the 
detector model.
When considering all nontiming distributions as constraining the timing distributions, the total constrained uncertainty is 4.1\% for 
\NCEOffTiming and 10.3\% for \NCPiOffTiming.

\subsection{\label{sec:fits}Fit method}
Two different confidence level limits are extracted from the data: (i) full nucleon, and (ii) electron. Each fit methodology is described below.

\subsubsection{Full nucleon}
For this fit the signal distributions were \NCEOff, \NCPiOff, \NCEOffTiming, and \NCPiOffTiming. 
The \CCQE, \NCE, and \NCPi distributions from neutrino and antineutrino modes, as well as \CCQEOff were used to constrain the 
systematic uncertainties and predicted beam-related backgrounds in the signal channels. 
The \CCQE and \NCE distributions are fitted as functions of $Q_{QE}^{2}$.
The $Q_{QE}^{2}$ for \NCE is obtained via 
\begin{equation}
Q_{QE}^{2} = 2m_{N}T_{N}^\text{reco}\label{eq:eq3},
\end{equation}
where $m_N$ is the effective mass of the nucleon and $T_N^\text{reco}$ is the reconstructed kinetic energy of the nucleon recoil. 
The \CCQE $Q_{QE}^{2}$ is obtained via
\begin{equation}
Q_{QE}^{2} = -m_\mu^2 + 2\EnuQE\left(E_\mu - \sqrt{E_\mu^2-m_\mu^2}\cos\theta_\mu\right)\label{eq:eq4},
\end{equation}
where
\begin{equation}
\EnuQE = \frac{2m_n^\prime E_\mu - \left[\left(m_n^\prime\right)^2 + m_\mu^2 - m_p^2\right]}{2\left[m_n^\prime - E_\mu + \sqrt{E_\mu^2-m_\mu^2}\cos\theta_\mu\right]}\label{eq:eq5},
\end{equation} 
and $E_\mu = T_\mu^\text{reco} + m_\mu$ is the total muon energy, and $m_p$, $m_n$ and $m_\mu$ are the masses of the proton, 
neutron and muon respectively. 
$m_n^\prime = m_n - E_B$ is the mass of the neutron minus the binding energy of carbon. 
A value of 34\MeV is used for $E_B$. Both equations arise from kinematic calculations assuming the incident nucleon is at rest. 
The \NCPi distributions, on the other hand, are fitted as a function of reconstructed $\pi^0$ momentum $p_{\pi^0}$. 
As already stated the \CCQEOff timing distribution was used to calibrate the simulated $T_\text{bunch}$, so it was not included in the 
dark matter fit.

During the fit, one normalization nuisance parameter was used for each mode of running, constrained by the integral of the 
corresponding \CCQE distribution. 
Two cross-section nuisance parameters were also used for each bin of the \NCE($Q_{QE}^{2}$) and \NCPi($p_{\pi^0}$) distributions, 
one for true neutrino and one for true antineutrino interactions. 
Neutrino and antineutrino interactions were considered separately because the neutrino/anti-neutrino interaction ratio is different between 
the three modes of running. 
In all, 23 nuisance parameters were used in the fit. 

Fake data sets were used in generating the confidence level limits. Fake data was generated by randomly sampling 
around the predicted distributions with a Gaussian distribution containing the full event covariance matrix. 
When a nonzero amount of dark matter is assumed, the dark matter distribution is added on top of the predicted distributions 
before generating the fake data set.

When setting the confidence level limits the nuisance parameters were held fixed to make the neutrino, antineutrino, and \CCQEOff 
distributions match the (fake) data being fitted. The nuisance parameters were held fixed because the signal distributions do not
affect the resulting nuisance parameters. It was also determined, with a small slice in the dark matter parameter space, that allowing the 
nuisance parameters to float did not alter the resulting confidence level limit, but did increase the computation time significantly. 
The dark matter signal rate (controlled by a scaling factor) was floated during the confidence level limit calculation.

In off-target mode 990 \CCQE, 1461 \NCE, and 148 \NCPi events were measured. 
After considering the constraints the predicted number of events is $1406\pm91$ and $135\pm15$ for \NCEOff and \NCPiOff respectively.
No significant excess was measured.

\subsubsection{Electron}
The signal distribution for this fit was defined as the events that pass \NCEE cuts with $\costhetae > 0.99$.
The fit was a binned extended maximum-likelihood fit in three dimensions, $\EvisE, \costhetae,\text{ and bunch time}$, with a single 
nuisance parameter to control the overall normalization of predicted neutrino events. 
The region of $0.9 < \costhetae < 0.99$ was the control region to constrain background events.
Because of the well-defined control region, data from neutrino and antineutrino modes were not used to constrain the prediction.
Two \NCEE events were measured in off-target mode.
After constraining the \NCEE background the predicted number of events is $2.4\pm 1.5$. 
In the signal region, zero events were measured with a constrained prediction of 0.4 events. 
Statistical error dominates the total error in the constrained prediction. 
No dark matter candidate events were measured.

Systematic uncertainties were not included in the fit as the predicted number of background events has a statistical relative uncertainty 
much greater than the predicted systematics, especially when considering some of the systematic uncertainties are constrained by the 
controlled region.
The normalization parameter is fixed during fitting so the data/fake data and null predictions are the same for the number of events in 
the control region.
When generating the fake data for the electron analysis, each bin is assumed to be independent with an underlying Poisson 
distribution with a mean equal to the predicted plus dark matter distribution.

\subsubsection{\label{sec:osc_events}Neutrino oscillation events in off-target mode}
\begin{figure}[htbp!]
\includegraphics[width=0.5\textwidth]{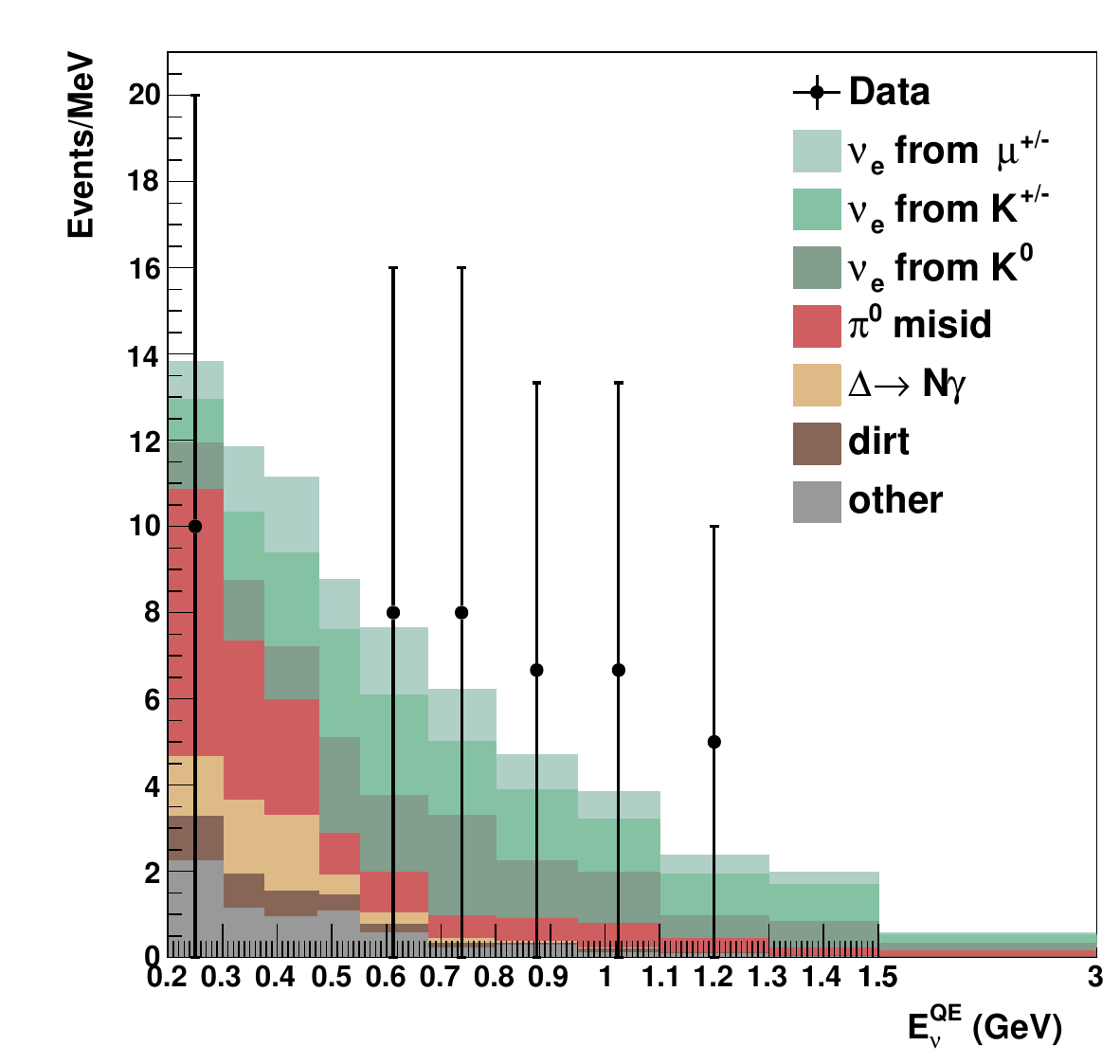}
  \caption{\label{fig:fig19}The \EnuQE distribution for events that pass the $\nu_e$ oscillation cuts. Data comes from off-target mode.}
\end{figure}

As previously stated about 40\% of the events that pass \NCEE cuts also pass neutrino oscillation cuts~\cite{Aguilar-Arevalo:2018gpe}. 
Figure~\ref{fig:fig19} shows the \EnuQE distribution [Eq.~(\ref{eq:eq5}) is used with the results from the electron track fit and $E_B = 0$] 
for off-target running. 
Simulation predicted 8.8 events assuming there are no oscillations. Six events were measured. All but one of the observed events were above 475\MeV.
Implications of this data are discussed in Sec.~\ref{sec:osc}.


\section{\label{sec:modelDep}Confidence Level Limits on Light Dark Matter Theory}
A fixed-target dark matter Monte Carlo simulator, \BdNMC, was used to simulate the energy and position distributions of the expected dark matter 
scattering signal in the \MB detector~\cite{deNiverville:2016rqh}. 
There are a number of production channels in fixed-target experiments, though often one will dominate for a given set of dark matter 
model parameters. 
For \MB, the decay of two pseudoscalar mesons, the $\pi^0$ and the $\eta$ were considered, as well as production through proton 
bremsstrahlung plus vector mixing up to $\mDMV = 1\GeVcc$.
The parameter values and equations used in the simulation were given in Ref.~\cite{deNiverville:2016rqh}.

The simulation loop begins by determining the maximum probability in
the angular momentum distribution of each production channel, as it is
not known analytically~\cite{deNiverville:2016rqh}. This maximum is used in an
acceptance-rejection algorithm to sample the angular momentum
distribution of each channel when generating dark matter trajectories.
The total number of dark matter particles expected from each
production channel is then calculated, and the output events are split
between these channels according to their fraction of the total dark
matter production rate.

For the case of pseudoscalar meson decays, meson 4-momenta and positions are generated in the \MB beam line by 
sampling an event list generated by the \textsc{BooNEG4Beam} simulations; see Sec.~\ref{sec:beamline}. 
For the case of proton bremsstrahlung, the dark matter is simulated to occur at the front of the beam dump.
  
  The simulation attempts a given dark matter scattering event for each dark matter trajectory from the previous step found to intersect 
  with the \MB detector. 
  Possible interactions are elastic-nucleon ($0\pi$), elastic-electron, and inelastic-nucleon producing a single pion ($1\pi^0$ if a $\pi^0$ is 
  produced, and $1\pi^\pm$ if a $\pi^\pm$ is produced). 
The neutrino detector simulation, discussed in Sec.~\ref{subsec:detector_sim}, was used to simulate the response of the detector. 
This simulation used neutrino events generated by \NUANCE and contained the nuclear model and all final-state interactions.
We define the weight of each neutrino simulated event as the ratio $N_\DMP\left(\omega\right)/N_\nu\left(\omega\right)$, where 
$N\left(\omega\right)$ is the number of true interactions as a function of energy transfer $\omega$ that generated the simulated event. 
$N_\DMP\left(\omega\right)$ is the number of interactions predicted by \BdNMC and $N_\nu\left(\omega\right)$ is the number of true 
interactions predicted by \NUANCE. Since $N_\nu\left(\omega\right)$ comes from the true distribution, which contains no nuclear model 
nor final-state interactions, the resulting reconstructed dark matter distribution contains the nuclear model and the model of final-state interactions 
that are in \NUANCE.
  
\begin{figure}[htpb!]
  \includegraphics[width=0.5\textwidth]{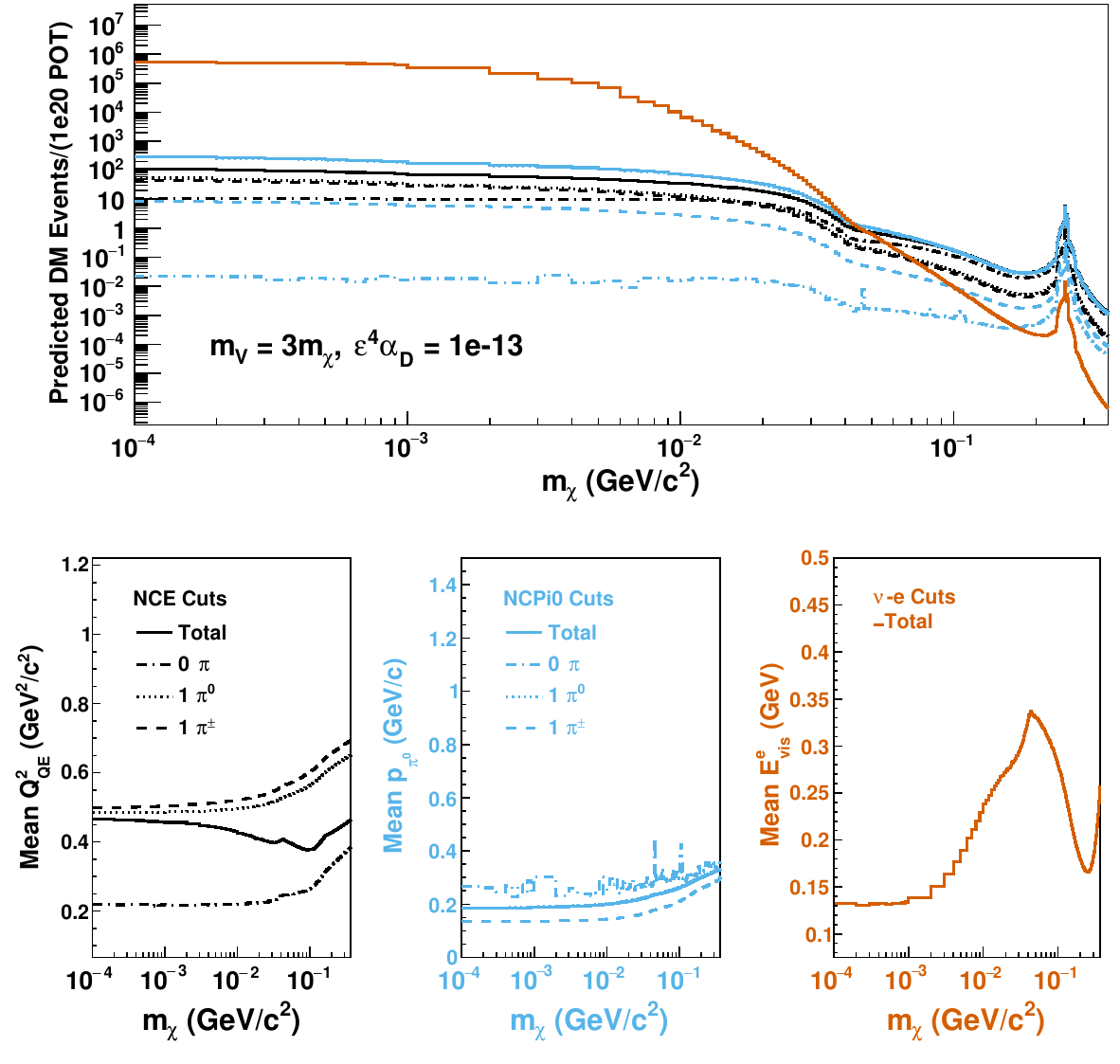}
  \caption{\label{fig:fig20} (Top) The integral number of events that pass \NCE, \NCPi, and \NCEE selection cuts as a 
  function of \mDMP. Predictions for the total, $0\pi$, $1\pi^0$, $1\pi^\pm$ are given by solid, dashed-dotted, dotted, and short-dashed lines 
  respectively. 
  The colors of the lines correspond with the bottom figures. 
  (Bottom left) The mean $Q_{QE}^{2}$ distribution for \NCE selection cuts. 
  (Bottom middle) The mean $p_\pi^0$ distribution for \NCPi selection cuts. 
  (Bottom right) The mean \EvisE distribution for \NCEE 
  selection cuts. All plots are functions of \mDMP with $\mDMV = 3\mDMP$ and $\DMScale = 1\times10^{-13}$.}
  \end{figure}

  Figure~\ref{fig:fig20} shows the number of events for \DMP scattering in the detector as well as the mean reconstructed 
  observables for $\mDMV = 3\mDMP$ and $\DMScale = 1\times10^{-13}$. 
At low masses the $1\pi$ dominates over $0\pi$ in overall rate for nucleon interactions. 
The $1\pi$ production dominate the \NCE distribution at higher $Q_{QE}^{2}$. 
Because of the separation of where $1\pi$ and $0\pi$ production dominates the \NCE distribution, and the efficiency of the \NCE 
selection cuts, \NCE provides significant constraint, along with \NCPi, on the lowmass region.
Dark matter scattering off electrons is predicted to dominate the overall rate at $\mDMP < 0.4\GeVcc$.

\begin{figure}
\subfloat[\NCE]{\includegraphics[width=0.24\textwidth]{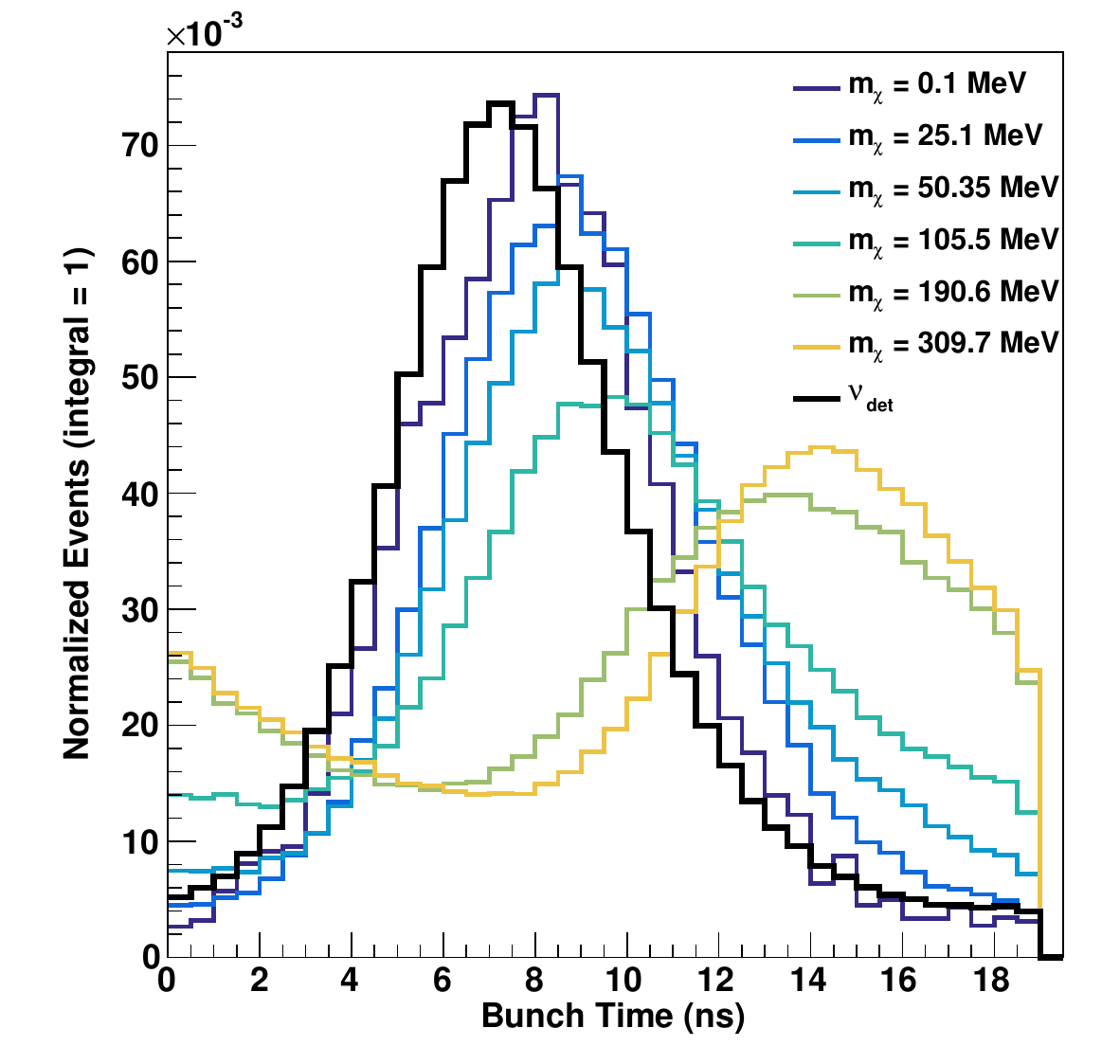}}
\subfloat[\NCPi]{\includegraphics[width=0.24\textwidth]{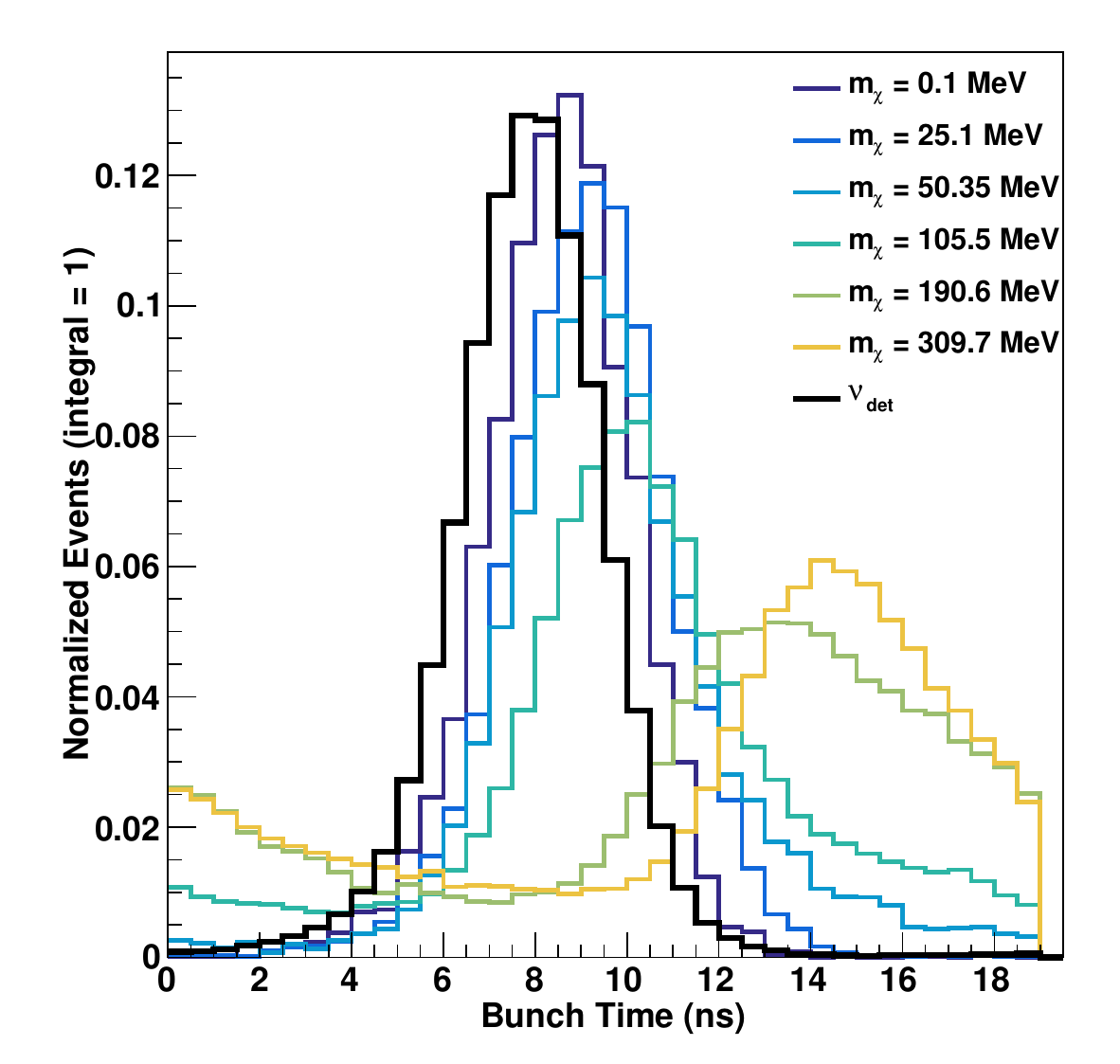}}
\caption{\label{fig:fig21}Comparison between signal interactions that pass (a) \NCE and (b) \NCPi cuts to different \DMS 
distributions as a function of bunch time. The neutrino interactions are for the neutrino mode while \DMS is for off-target mode (see text).}
\end{figure}

Figure~\ref{fig:fig21} compares the bunch-time distribution for various combinations of \mDMV, \mDMP to the neutrino distribution 
used for the candidate signal events that pass \NCE and \NCPi selection cuts.
The neutrino distribution is the predicted distribution for \nuModeLong running, while the \DMS distributions are for off-target mode. 
The difference between the neutrino distribution and that of the lightest \DMS mass represents the difference between \nuModeLong 
and off-target running, which is consistent with the average time difference between the dark matter and the neutrino to reach steel beam dump. 
The sensitivity for heavier dark matter masses is improved when using timing.
 
\begin{figure}[htbp]
  \subfloat[\label{fig:fig22a}Full Nucleon + Timing]{\includegraphics[width=0.5\textwidth]{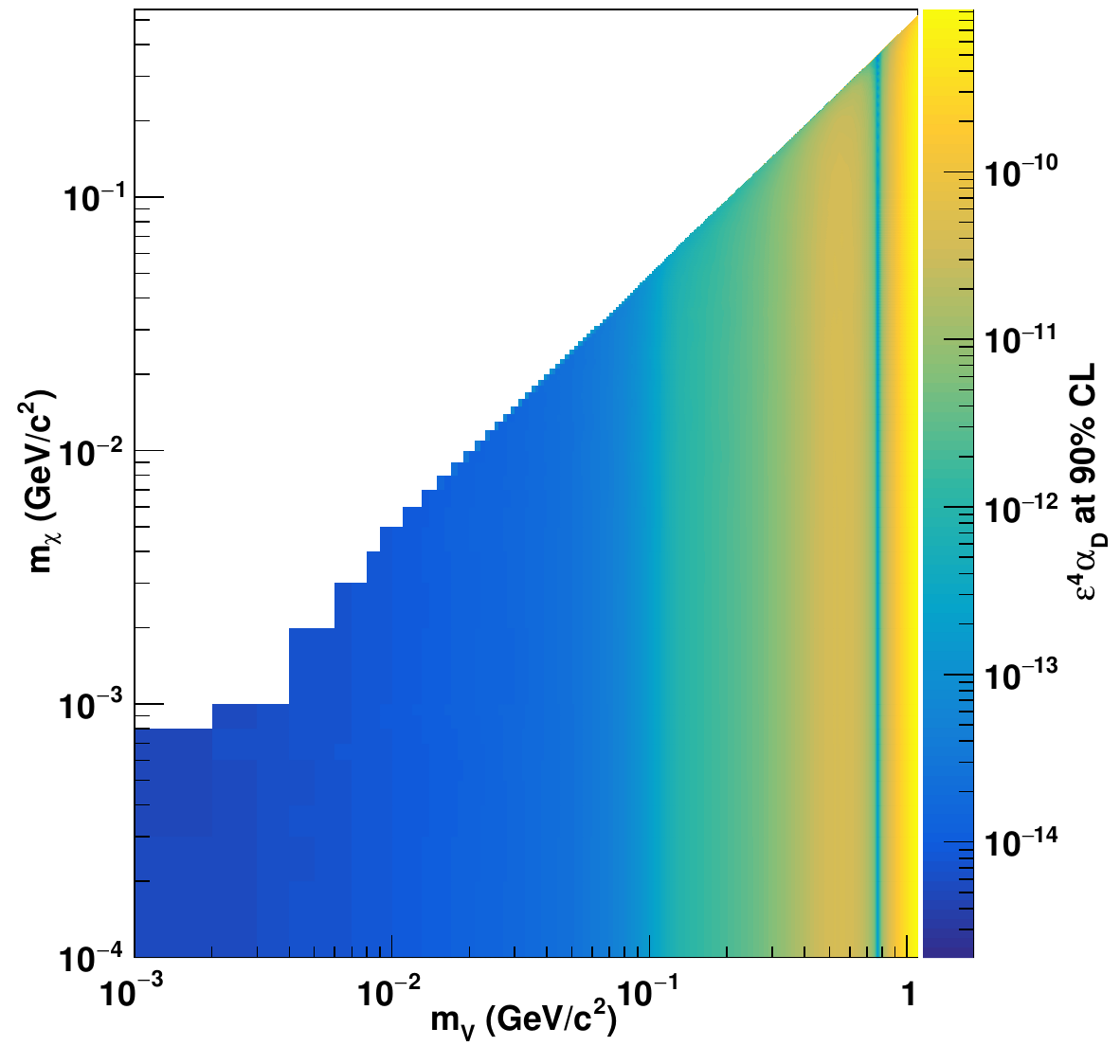}}\\
  \subfloat[\label{fig:fig22b}Electron+Timing]{\includegraphics[width=0.5\textwidth]{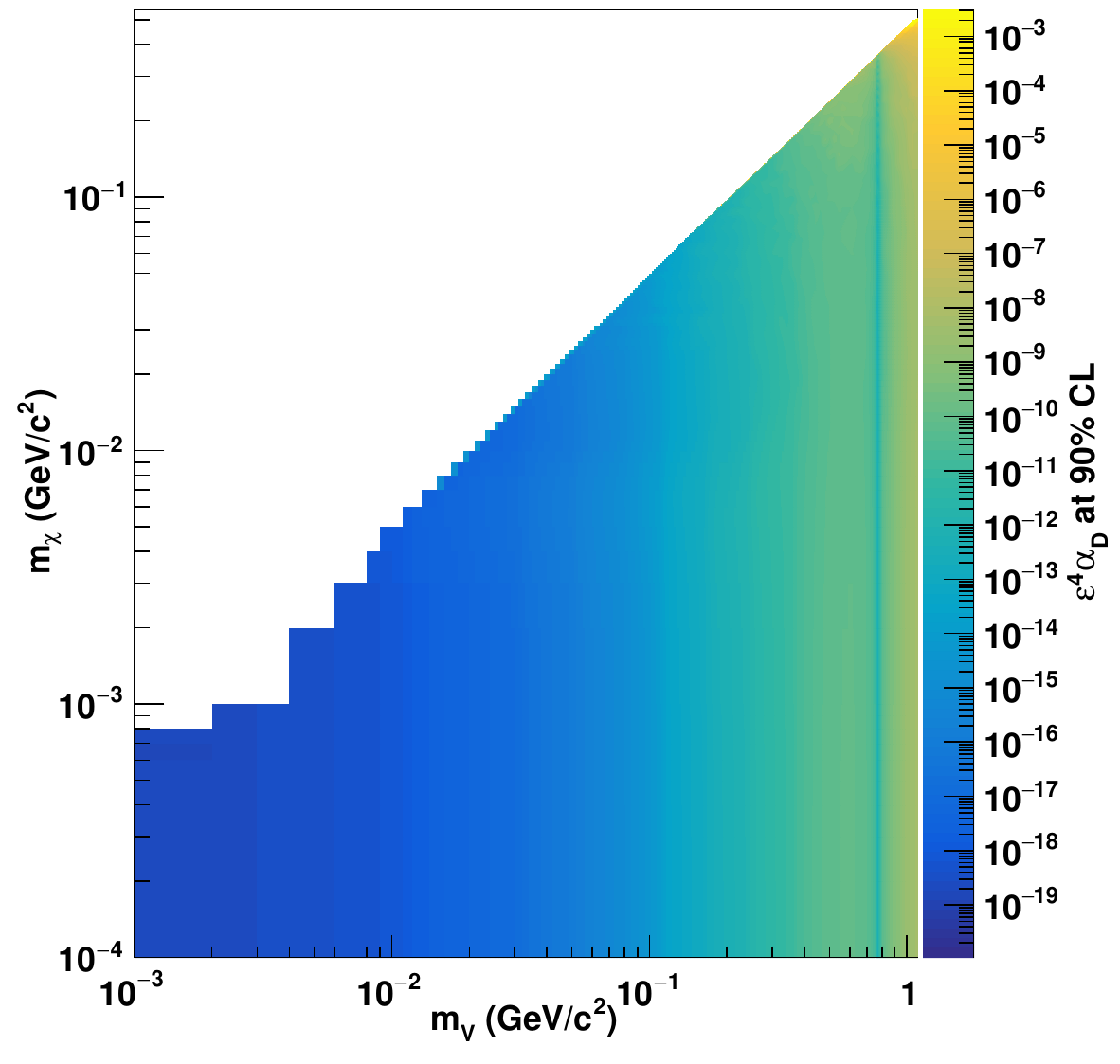}}
  \caption{\label{fig:fig22}The 90\% confidence level limit for (a) full nucleon with timing and (b) electron with timing on $\kineticMixing^4\darkFineStructure$ for various combinations of \mDMV and \mDMP using the vector portal dark matter model.}
\end{figure}

Using the results from \BdNMC and the frequentist confidence level method developed for the \MB oscillation 
analysis~\cite{Aguilar-Arevalo:2013pmq}, 90\% confidence level limits were calculated for different combinations of \mDMV and \mDMP as a 
function of $\kineticMixing^4\darkFineStructure$.
The frequentist approach used fake data and various fits to fake data to generate the effective degrees of freedom given a predicted signal. 
Each combination of \mDMV and \mDMP were treated independently, and because only on-shell decay was considered (see Sec.~\ref{sec:theory}), $\kineticMixing^4\darkFineStructure$ controls only the normalization of the predicted \DMS signal. 
Figure~\ref{fig:fig22} gives the 90\% confidence level limits on $\kineticMixing^4\darkFineStructure$ as a function of $\mDMV\text{ and }\mDMP$ for both the full 
nucleon and electron fits when including timing.
The best limit, in the tested parameter space, was set at $\mDMV = 0.3\MeVcc$, $\mDMP = 0.1\MeVcc$ with 
$\kineticMixing^4\darkFineStructure = 3.9\times10^{-15}$ for the full nucleon fit and $\mDMV = 0.5\MeVcc$, $\mDMP = 0.2\MeVcc$ with 
$\kineticMixing^4\darkFineStructure = 1.3\times10^{-19}$ for the electron fit. 

\begin{figure}[htbp!]
\includegraphics[width=0.5\textwidth]{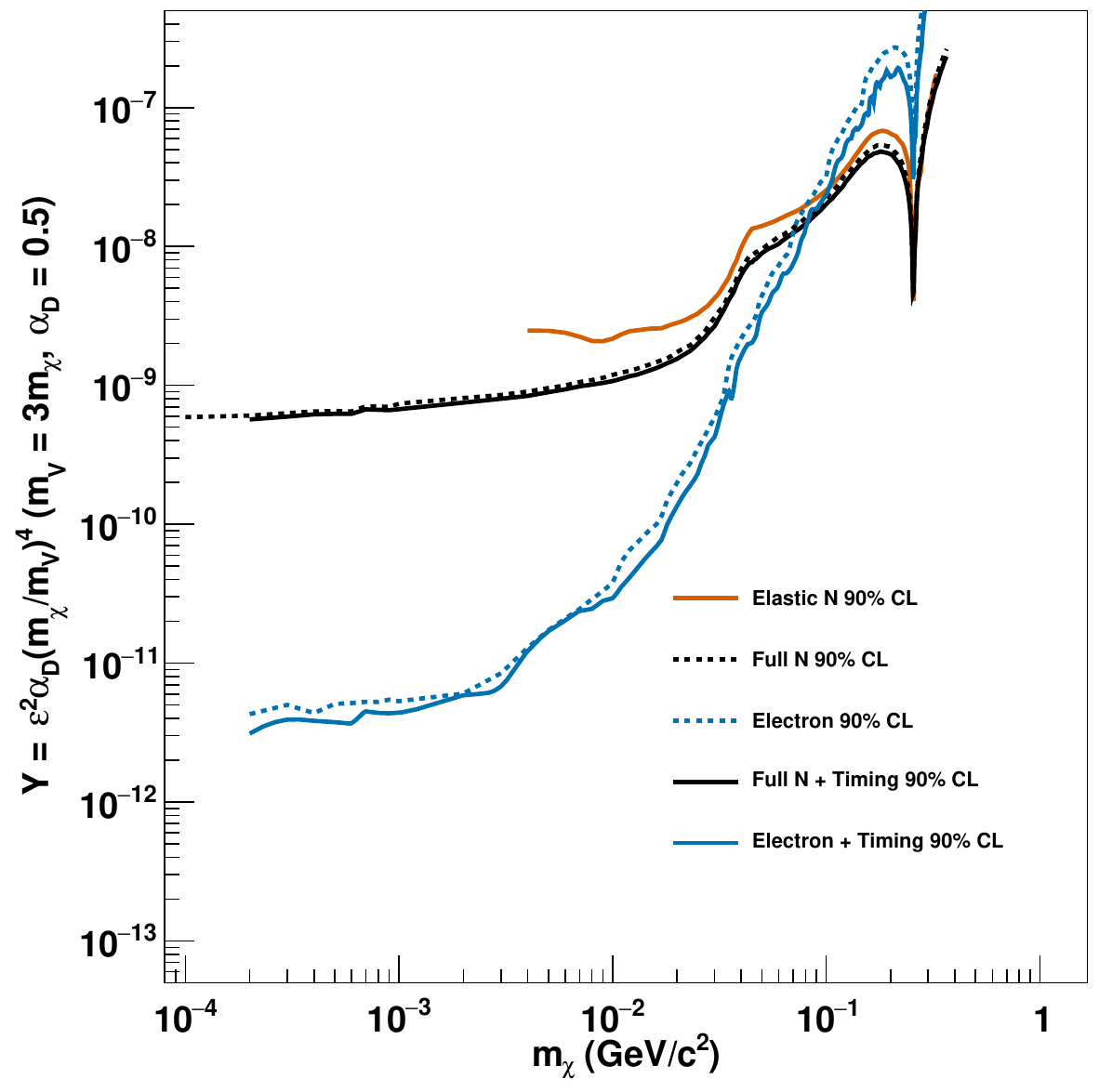}
\caption{\label{fig:fig23}Comparing the full nucleon and electron confidence level results to the elastic nucleon results from 
Ref.~\cite{Aguilar-Arevalo:2017mqx}. Also shown is the result when including timing (solid lines) compared to that when not including the 
timing (dashed lines).}
\end{figure}

Figure~\ref{fig:fig23} compares the confidence level results in this paper to the elastic-nucleon results~\cite{Aguilar-Arevalo:2017mqx} 
for the \DMS parameter slice $\mDMV = 3\mDMP$ and $\darkFineStructure = 0.5$, where 
$Y = \kineticMixing^2\darkFineStructure\left(\mDMP/\mDMV\right)^4$ is a dimensionless parameter that controls the dark matter 
annihilation cross section and in turn the thermal relic abundance.
Also shown are the confidence level limits when timing is not included.
Including the $1\pi$ \DMP interaction channels improves the confidence level from Ref.~\cite{Aguilar-Arevalo:2017mqx} at low masses while 
including timing improves the confidence level limits at high masses in $\kineticMixing^4\darkFineStructure$ up to a factor of 1.5 for the full 
nucleon fit and 4.7 for the electron fit. 
For the variable $Y$ this corresponds to improvements in the confidence level limits by 1.2 and 2.2 for the full nucleon and electron fits respectively. 
The electron fit gives more restrictive limits at lower masses compared to the full nucleon fit.


\section{\label{sec:results}Results and Discussion}
The following is a discussion on the implications of the results presented above. 
A comparison of the full nucleon and electron dark matter analyses with current limits will be discussed, 
followed by the implications of not seeing an excess in the neutrino oscillation sample.
The section will conclude with the implications of running with a proposed dedicated ``beam-dump'' target. 

\subsection{Limits on sub-GeV dark matter}
\MB has improved upon the results published in Ref.~\cite{Aguilar-Arevalo:2017mqx} through dedicated searches for $\pi^0$ 
production and elastic scattering initiated by dark matter particles produced in a proton beam dump.
The dark matter search built upon a rich history of cross section and oscillation analyses already published by the \MB Collaboration.
The full nucleon dark matter analysis leveraged neutrino and anti-neutrino data sets, as well as the \CCQEOff distribution to 
constrain systematic uncertainties.
Both the full nucleon and electron analyses utilized the use of the BNB bunch structure to set stronger limits on heavier \mDMP.

\begin{figure*}[htbp!]
  \subfloat[\label{fig:fig24a}vector portal]{\includegraphics[width=0.5\textwidth]{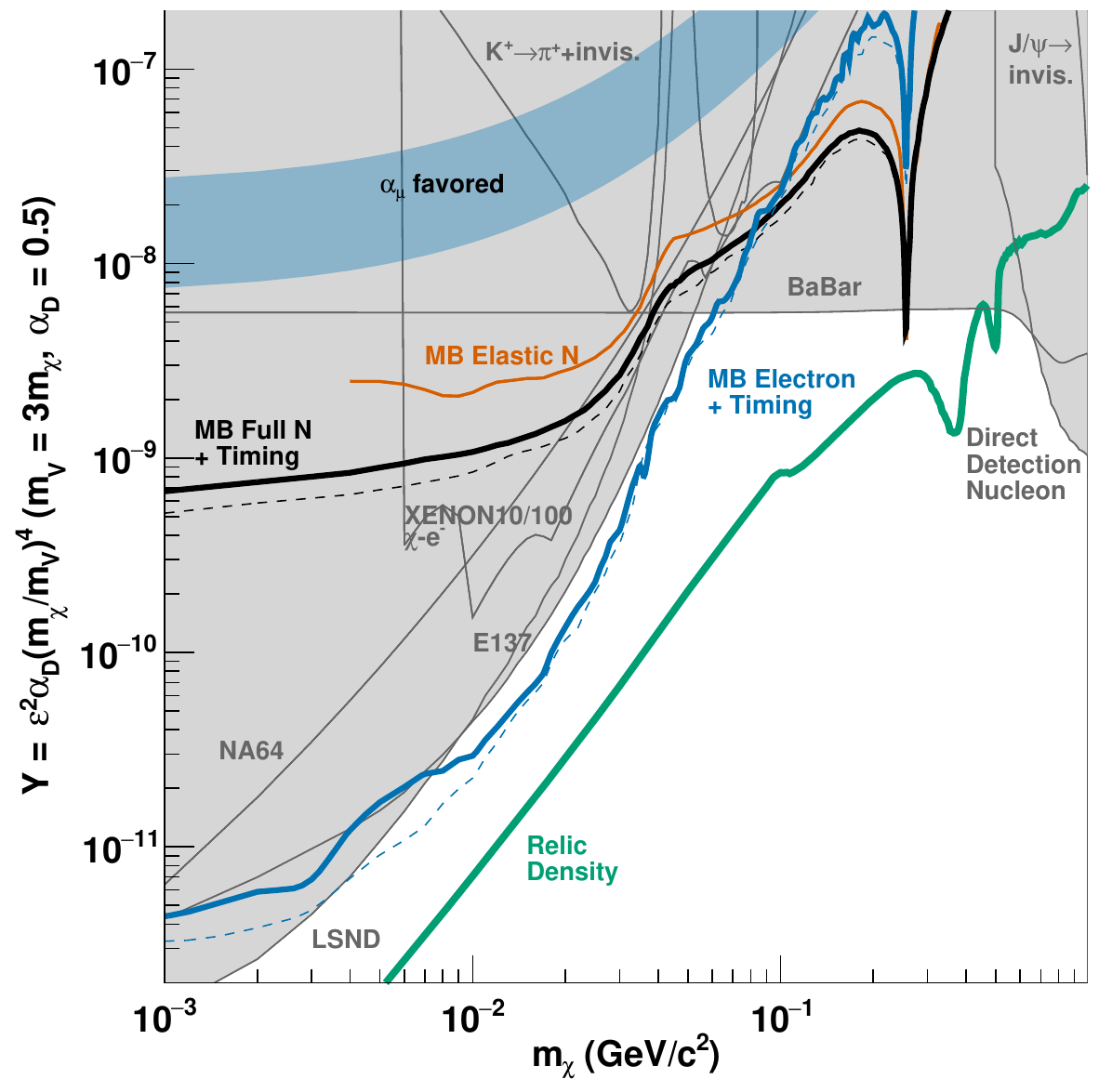}}
   \subfloat[\label{fig:fig24b}leptophobic]{\includegraphics[width=0.5\textwidth]{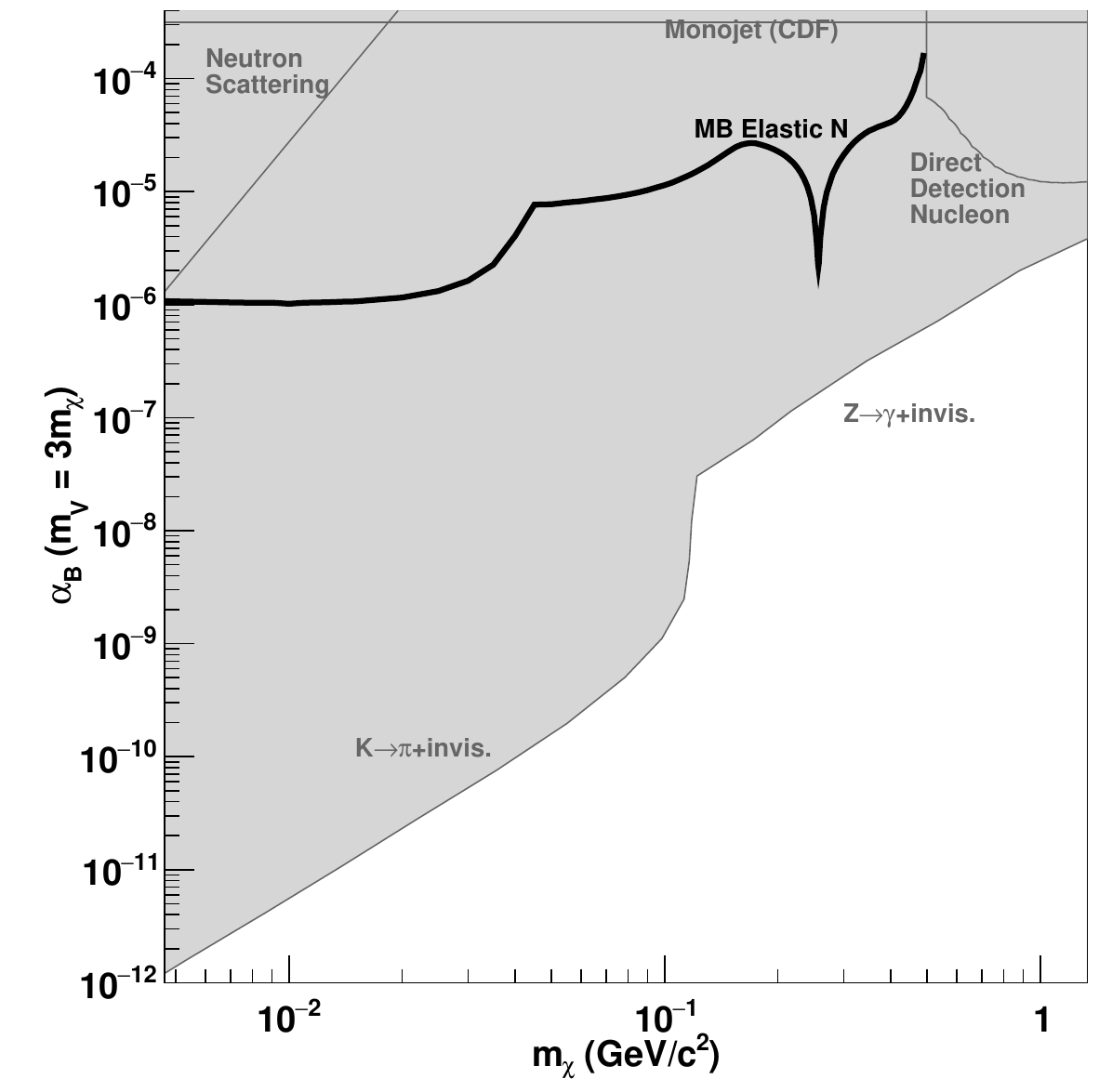}}
  \caption{\label{fig:fig24}Comparison of the \MB confidence level limits (solid lines), and sensitivities (dashed lines) to other experiments for 
  (a) $Y$ as a function of \mDMP assuming \darkFineStructure = 0.5 and \mDMV = 3\mDMP and (b) in the leptophobic dark 
  matter model with \mDMV = 3\mDMP.
   An explanation of vector portal limits lines was given in 
   Refs.~\cite{Batell:2014mga,Banerjee:2017hhz,Essig:2017kqs,deNiverville:2016rqh,Lees:2014xha}.
 An explanation of the leptophobic limit lines was given in Refs.~\cite{Batell:2014yra,Dror:2017ehi,Dror:2017nsg}. 
   }
\end{figure*}

\begin{figure*}[htbp]
   \subfloat[\label{fig:fig25a}vector portal ($\darkFineStructure=0.1, \mDMV = 3\mDMP$)]{
   \includegraphics[width=0.5\textwidth]{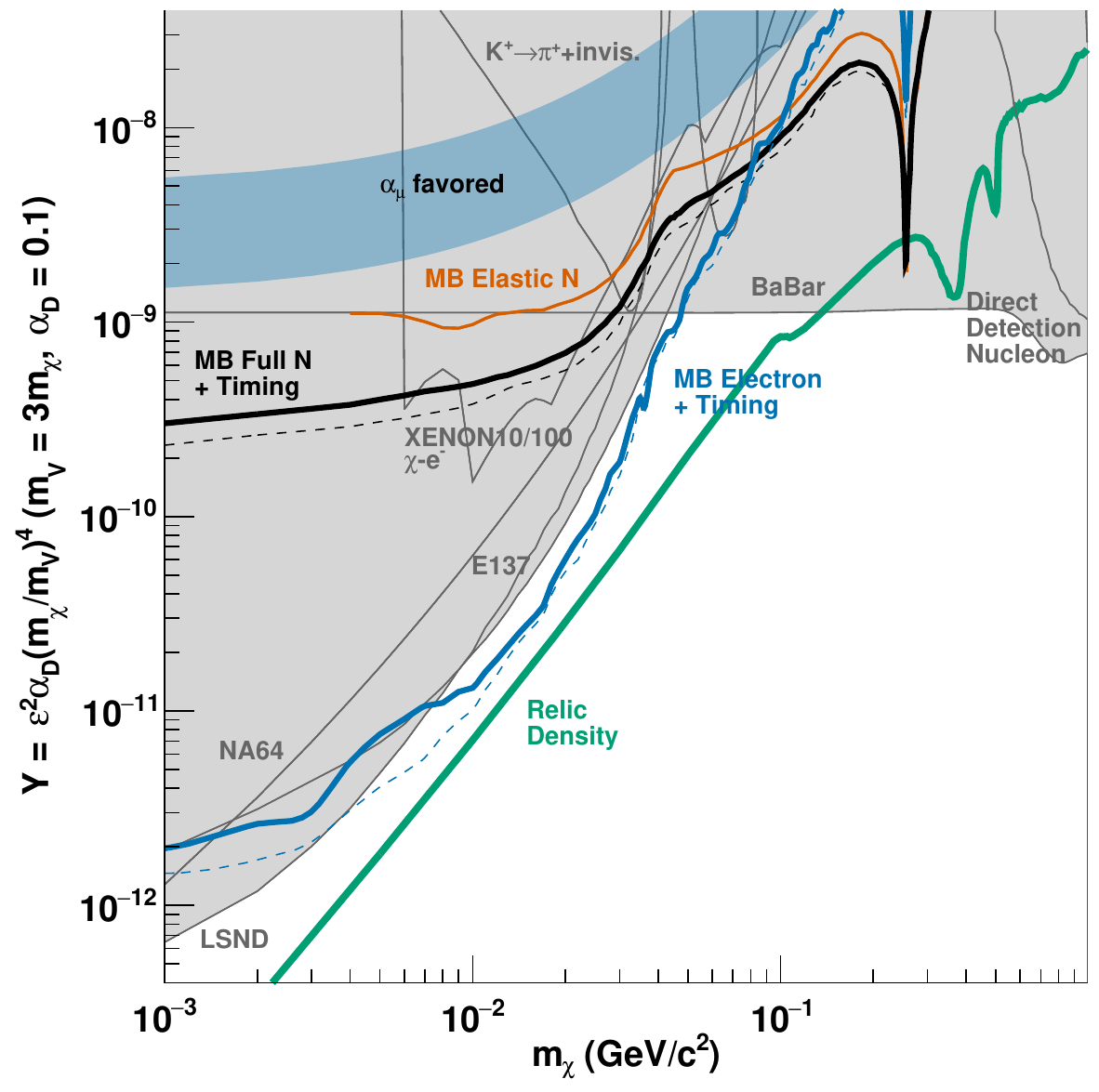}}
  \subfloat[\label{fig:fig25b}vector portal ($\darkFineStructure=0.5, \mDMV = 7\mDMP$)]{
  \includegraphics[width=0.5\textwidth]{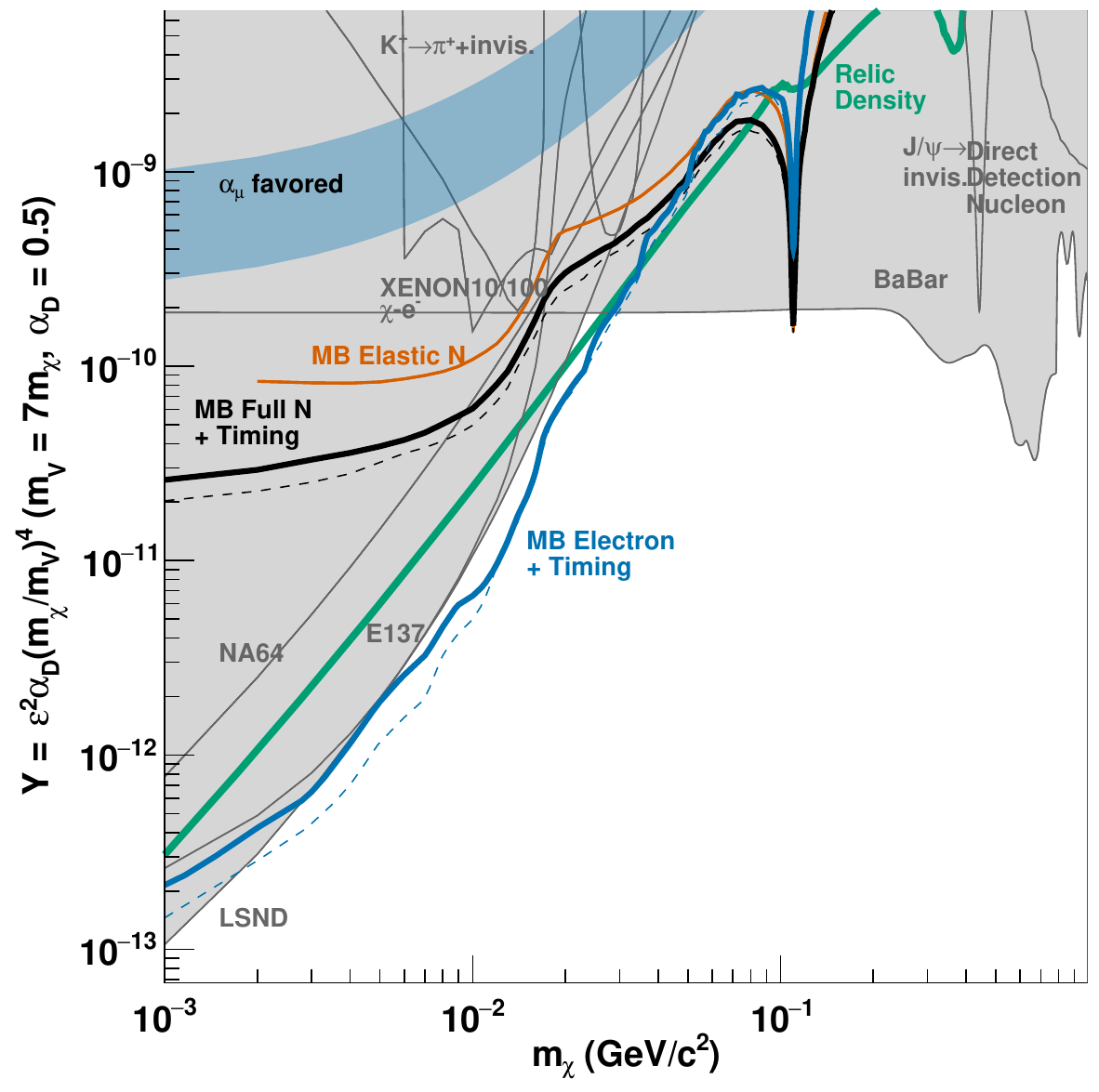}}
 \caption{\label{fig:fig25} 90\% confidence level in the vector portal dark matter model with 
 (a) $Y$ as a function of \mDMP assuming \darkFineStructure = 0.1 and \mDMV = 3\mDMP and (b) \darkFineStructure = 0.1 and 
 \mDMV = 7\mDMP. An explanation of the limit lines was given in 
 Refs.~\cite{Batell:2014mga,Banerjee:2017hhz,Essig:2017kqs,deNiverville:2016rqh,Lees:2014xha}}
 \end{figure*}

Figures~\ref{fig:fig24a},~\ref{fig:fig25a}, and~\ref{fig:fig25b} show three example 
projections of the limits in Fig.~\ref{fig:fig22} to the \mDMP-$Y$ plane.
The chosen projections are standard but are not the only ones possible. 
The differences between the three slices are due to different assumptions about $\darkFineStructure$ and the relationship 
between \mDMP and \mDMV. 

Two different relationships between \mDMP and \mDMV are shown to demonstrate how the contours in the \mDMP-$Y$ 
plane change as a function of model parameters. 
The relic density contour (green) indicates the \mDMP, $Y$ values where the model with a complex scalar \DMS particle 
predicts a dark matter abundance that is in accord with observations.
As the ratio $\mDMV/\mDMP$ increases, the constraints on the thermal relic target become more stringent and can be 
ruled out over the full parameter space; see Fig.~\ref{fig:fig25b}. 
Furthermore, as the ratio $\mDMV/\mDMP$ increases, other kinds of dark matter scenarios, e.g. asymmetric fermionic 
dark matter, can be probed by \MB and other current experiments~\cite{Battaglieri:2017aum}. 
At smaller $\mDMV/\mDMP$ ratios there is still a wide region of parameter space in the complex scalar \DMS model that 
can satisfy the relic density requirement; see Figs.~\ref{fig:fig24a} and~\ref{fig:fig25a}. 

For the vector portal model, \MB excludes the muon {\it g--2} favored region, and some regions where this model satisfies 
the relic density in the parameter space tested. 
\MB also excludes previously untested parameter space, especially in the electron channel. 
For the leptophobic dark matter scenario, inelastic neutral pion production has not been studied in the literature.
Therefore, the nucleon elastic results from Ref.~\cite{Aguilar-Arevalo:2017mqx} were used to place conservative limits 
on this scenario. 
The result can be found in Fig.~\ref{fig:fig24b}.

\subsection{\label{sec:osc}\MB Neutrino oscillations excess}
\MB has recently doubled the amount of \nuModeLong POT~\cite{Aguilar-Arevalo:2018gpe}. 
The reported neutrino plus antineutrino oscillation excess is $460.5\pm99.0$ for a combined $24.11\times10^{20}$\,POT. 
If this excess were due to a process that is occurring in the beam dump, such as dark matter production, instead of neutrino-related 
processes, the predicted excess would scale with the amount of POT collected.

An example process that would scale solely by POT would be the production of a dark mediator through neutral meson 
decay or proton bremsstrahlung which would then decay into two dark matter particles.
One of the dark matter particles would then decay in the detector producing a lepton-antilepton pair with low invariant mass. 
A potential dark matter model that could be extended to fit such a description can be found in Refs.~\cite{Izaguirre:2017bqb,Jordan:2018gcd}.

The predicted off-target excess, under this assumption, is $35.5\pm7.6$, whereas the measured excess is -2.8 events integrated over 
$200 \le \EnuQE < 1250\MeV$, see Sec.~\ref{sec:osc_events}. 
Assuming Gaussian errors, the measured off-target sample of events that pass oscillation cuts is inconsistent, at $4.6\sigma$, with a
process that predicts all of the oscillation excess scales with the collected POT independent of the beam configuration.

\subsection{Proposed dedicated ``beam-dump'' target}
\MB has shown that a neutrino experiment can search for fixed-target accelerator-produced dark matter scattering for different 
production and interaction channels. 
Most of the neutrino backgrounds came from proton interactions in the air and scraping of the target. 
To further reduce the neutrino background a dedicated ``beam-dump'' target is needed. 
A simulation of a steel beam dump target positioned where the neutrino target/horn are located, effectively removing the 
decay pipe, indicates the decrease of the \CCQEOff event rate by a factor of 20.
The \NCPi and \NCEE sensitivities would increase the most with this reduction in the beam-related backgrounds. 
For example, a total of five events are predicted to pass \NCPi selection cuts for $1.86\times10^{20}\,\mathrm{POT}$ compared to 
the 148 measured in this analysis.
The reduction for \NCPi is larger than \NCE or \CCQE because more energetic neutrinos are required to generate \NCPi events.

A dedicated ``beam-dump'' target would also decisively test theories that predict the oscillation excess scales as POT. With a dedicated 
``beam-dump'' target almost no events are expected to pass oscillation cuts. 
An upgrade is being considered that would add a secondary ''beam-dump'' target to the 
BNB~\footnote{See talk by R.G. Van de Water, \url{https://indico.fnal.gov/event/15726/session/3/material/0/0.pdf}}. 
The addition of the second target would allow simultaneous running, on a pulse-by-pulse basis, of protons hitting the neutrino and 
``beam-dump'' targets. 
This would increase the physics output of the Short-Baseline Neutrino Program~\cite{Antonello:2015lea} at Fermilab.

\section{Acknowledgements}
This  work  was  supported  by  the  U.S. Department of Energy;  the U.S. National Science Foundation;  Los  Alamos  National  Laboratory; 
the Science and Technology Facilities Council, UK; Consejo Nacional de Ciencia y Tecnologa, Mexico.  We thank the Fermilab Accelerator Division 
for the work to reconfigure, operate, and understand the off-target beam.  Fermilab is operated by Fermi Research Alliance, LLC under 
Contract No.  De-AC02-07CH11359 with the US DOE. We also thank Los Alamos National Laboratory for LDRD funding.

\bibliography{references}

\end{document}